\documentclass[review]{elsarticle}
\usepackage{hyperref}
\usepackage {amssymb}
\usepackage {amsmath,amsthm}
\usepackage {a4}
\usepackage {graphicx}
\newcommand {\e} {\varepsilon}
\usepackage{color} 

\DeclareMathOperator*{\supp}{supp}
\DeclareMathOperator*{\sgn}{sgn}

\newcommand \beq{\begin{equation}}
\newcommand \eeq{\end{equation}}
\newcommand \sskip{\smallskip\noindent}
\newtheorem{remark}{Remark}[]

\newcommand{\beginsupplement}{%
        \setcounter{table}{0}
        \renewcommand{\thetable}{S\arabic{table}}%
        \setcounter{figure}{0}
        \renewcommand{\thefigure}{S\arabic{figure}}%
     }

\journal{}

\begin{document}

\begin{frontmatter}

\title{Trade-offs between chemotaxis and proliferation shape the phenotypic structuring of invading waves}

\author{Tommaso Lorenzi}
\address{Department of Mathematical Sciences ``G. L. Lagrange'', Dipartimento di Eccellenza 2018-2022, Politecnico di Torino, 10129 Torino, Italy (tommaso.lorenzi@polito.it)}

\author{Kevin J Painter}
\address{Inter-university Department of Regional and Urban Studies and Planning, Politecnico di Torino, 10129 Torino, Italy (kevin.painter@polito.it)}




\begin{abstract}
Chemotaxis-driven invasions have been proposed across a broad spectrum of biological processes, from cancer to ecology. The influential system of equations introduced by Keller and Segel has proven a popular choice in the modelling of such phenomena, but in its original form restricts to a homogeneous population. To account for the possibility of phenotypic heterogeneity, we extend to the case of a population continuously structured across space, time and phenotype, where the latter determines variation in chemotactic responsiveness, proliferation rate, and the level of chemical environment modulation. The extended model considered here comprises a non-local partial differential equation for the local phenotype distribution of cells which is coupled, through an integral term, with a differential equation for the concentration of an attractant, which is sensed and degraded by the cells. In the framework of this model, we concentrate on a chemotaxis/proliferation trade-off scenario, where the cell phenotypes span a spectrum of states from highly-chemotactic but minimally-proliferative to minimally-chemotactic but highly-proliferative. Using a combination of numerical simulation and formal asymptotic analysis, we explore the properties of travelling-wave solutions. The results of our study demonstrate how incorporating phenotypic heterogeneity may lead to a highly-structured wave profile, where cells in different phenotypic states dominate different spatial positions across the invading wave, and clarify how the phenotypic structuring of the wave can be shaped by trade-offs between chemotaxis and proliferation.
\end{abstract}

\begin{keyword}
Chemotaxis invasion \sep Phenotypic heterogeneity \sep Travelling waves \sep Non-local partial differential equations
\MSC[2020] 92C17 \sep 35Q92 \sep 35R09 \sep 35B40 \sep 35C07 
\end{keyword}

\end{frontmatter}

%\linenumbers

%\linenumbers

\section{Introduction}
More than half a century ago, seminal studies by Adler~\cite{adler1966} demonstrated that the placement of a small population of {\em E. coli} bacteria in a suitable nutrient environment led to sustained high-density travelling bands or rings that spread outwards. Critical to this phenomenon was the chemotactic behaviour of the bacteria, which by consuming the nutrient establishes an attractant gradient that persistently propels the population from nutrient-poor to nutrient-rich regions. In subsequent years, chemotaxis and other taxis-like motility behaviours have been implicated in a broad variety cell migration and invasion processes, including embryonic development~\cite{giniunaite2020,szabo2018} and cancer invasion~\cite{condeelis2005,stuelten2018}. 

\sskip
The work of Keller and Segel is indelibly linked to the modelling of chemotaxis phenomena and one of a triumvirate of seminal studies of the early 1970s specifically addressed the travelling-band behaviour observed in Adler's experiments~\cite{keller1971a}. The chemotaxis model of Keller and Segel takes the form of a system of advection-reaction-diffusion equations, with chemotaxis modelled as an advective process that drives the density of individuals up (or down) the concentration gradient of some chemical attractant (or repellent). A very large modelling literature has since emerged, applied to a wide spectrum of cellular, ecological and social phenomena~\cite{painter2019}, and explicit analytical investigations into travelling-wave dynamics have received considerable attention~\cite{wang2013}.  

\sskip
One assumption of Keller and Segel's original model, though, is that the population is homogeneous. In short, each individual is assumed to respond and react in the same manner: the same potential to divide, the same motility response, the same rate of nutrient consumption {\em etc}. Such homogeneity is rarely present in cellular systems, where significant variation can be present within a population of superficially identical cells; for example, a spectrum of gene expression profiles that weight a cell towards one form of behaviour (say, growth) over another (say, motility). 

\sskip
Interindividual variability in {\em phenotypes} (i.e. observable physical or biochemical characteristics) can be incorporated into mathematical models for spatio-temporal population dynamics by introducing a continuous structuring variable that describes the phenotypic state of every individual and then defining a suitable evolution equation for the function that represents the individual distribution in both physical and phenotype spaces (i.e. the population density function). Mathematical models of this type -- whereby the evolution equation for the population density function takes the form of a non-local reaction-diffusion equation~\cite{arnold2012existence,benichou2012front,berestycki2015existence,bouin2014travelling,bouin2012invasion,bouin2017super,turanova2015model} or a non-local advection-reaction-diffusion equation~\cite{lorenzi2021} -- have received considerable attention from the mathematical community over the last few years, and have been increasingly used as theoretical tools to dissect out the mechanisms which underpin the spatial spread and phenotypic evolution of populations with heterogeneous motility responses.

\sskip
In this paper, we present a mathematical model for chemotaxis in which a continuous structuring variable captures intercellular variability in proliferative potential and chemotactic sensitivity. The model comprises a non-local advection-reaction-diffusion equation for the cell population density function that is coupled, through an integral term, with a differential equation for the concentration of an attractant, which is sensed and degraded by the cells. Compared to related models considered in previous studies~\cite{arnold2012existence,benichou2012front,berestycki2015existence,bouin2014travelling,bouin2012invasion,bouin2017super,turanova2015model,lorenzi2021}, our model takes into account chemotactic movement alongside nonlinear dynamical interactions between individuals and the attractant, which may induce feedback mechanisms regulating population growth. This brings about richer spatio-temporal dynamics, including the emergence of both travelling fronts and travelling pulses with compact support.

\sskip
In the framework of this model, we investigate the impact of chemotaxis/proliferation trade-offs on the spatial eco-evolutionary dynamics of the cells. We consider scenarios in which cell proliferation is independent from the attractant and where the attractant is viewed in the light of a nutrient that fuels cell proliferation. The latter is a natural enough scenario in a number of biological contexts, for example nutrients including glucose and other energy sources for {\em E. coli} bacteria \cite{adler1966} or molecular growth factors that trigger both proliferation and movement during developmental, physiological and pathological processes; specific examples of the latter include interleukin-8 in cancer growth/invasion \cite{waugh2008} or PDGF$\beta$ during wound repair \cite{wang2019}. Furthermore, we consider cases in which attractant degradation is not directly linked to the proliferation rate of cells and where, assuming the attractant acts as a nutrient, higher rates of proliferation demand greater consumption of the attractant. Finally, we explore how the spatial eco-evolutionary dynamics of the cells may be affected by environment-induced phenotypic changes, regulated by a nutrient-type attractant. In particular, we suppose that cells in poor-nutrient conditions become more exploratory in order to seek out better regions, and cells in a good-nutrient environment remain relatively motionless in order to feed and proliferate.

\sskip
The remainder of the paper is organised as follows. In Section~\ref{Sec:Model}, we describe the model and the main underlying assumptions. In Section~\ref{Sec:Analysis} we carry out formal asymptotic analysis of the model equations. The results obtained are subsequently integrated with the results of numerical simulations in Section~\ref{Sec:Numerics}. In Section~\ref{Sec:Discussion}, we briefly explain how these mathematical results may shed light on the mechanisms that govern the spatial spread and phenotypic evolution of heterogeneous cell populations in the presence of chemotactic cues. Moreover, we provide a brief overview of possible research perspectives.

\section{Statement of the problem} 
\label{Sec:Model}
\subsection{Mathematical model}
We consider a mathematical model for the dynamics of a growing population of cells which sense the spatial gradient of an attractant. The cell population is structured by a variable $y \in [0,Y] \subset \mathbb{R}_+$, with $Y>0$, which represents the phenotypic state of each cell and takes into account intra-population heterogeneity in the cell proliferation rate and the cell chemotactic sensitivity (\emph{e.g.} the variable $y$ could represent the level of expression of a gene that regulates both cell proliferation and cell sensitivity to chemotactic cues). Focussing on a one-dimensional spatial domain scenario, the local cell phenotype distribution at position $x \in \mathbb{R}$ and time $t \in [0,\infty)$ is modelled by the cell population density function $n(x,y,t)$ while the concentration of the attractant is described by the function $S(x,t)$. The evolution of $n(x,y,t)$ is governed by the following non-local advection-reaction-diffusion equation
\beq
 \label{eq:PDEn}
 \begin{cases}
\displaystyle{\partial_t n + \chi(y) \, \partial_x \left(n  \, \partial_x S \right) + \partial_y \left(\phi \left(y,S\right) n \right) = R(y,\rho,S) \, n + a \, \partial^2_{xx} n +  b \, \partial^2_{yy} n,}
\\\\
\displaystyle{\rho(x,t) := \int_{0}^Y n(x,y,t) \, {\rm d}y,}
\end{cases}
\,
(x,y) \in \mathbb{R} \times (0,Y),
\eeq
coupled with the following differential equation for $S(x,t)$ 
\beq
 \label{eq:PDES}
\partial_t S = - \int_0^Y \kappa(y,S) \, n(x,y,t) \, {\rm d}y, \quad x \in \mathbb{R},
\eeq
which relies on the assumption that the attractant is solely consumed by the cells (i.e. diffusion and production of the attractant are neglected). We remark that within the spirit of a general model not tailored towards a specific application, this is primarily a model simplification that aids analytical calculations -- we refer to Section~\ref{Sec:Discussion} for a discussion regarding the generalisation to other scenarios. Here, \eqref{eq:PDEn} is subject to zero-flux boundary conditions at $y=0$ and $y=Y$.

\paragraph{Chemotactic movement} The second term on the left-hand side of~\eqref{eq:PDEn} represents the rate of change of the cell population density due to chemotactic movement (i.e. cell movement up the gradient of the attractant). The function $\chi(y)$ models the chemotactic sensitivity of cells in the phenotypic state $y$. Without loss of generality, we consider the case where higher values of $y$ correlate with higher cell chemotactic sensitivity and, therefore, we let the function $\chi(y)$ satisfy the following assumptions
\beq
 \label{ass:chi}
\chi(0) \geq 0, \quad \dfrac{{\rm d} \chi(y)}{{\rm d} y} > 0 \; \text{ for } y \in (0,Y).
\eeq

\paragraph{Cell proliferation and death} The first term on the right-hand side of~\eqref{eq:PDEn} represents the rate of change of the population density due to cell proliferation and death. The function $R(y,\rho(x,t),S(x,t))$ models the fitness (i.e. the net proliferation rate) of cells in the phenotypic state $y$ at time $t$ and position $x$ under the local environmental conditions given by the total cell density, $\rho(x,t)$, and the concentration of attractant, $S(x,t)$. We consider scenarios in which cell proliferation is independent from the attractant and where the attractant is viewed in the light of a nutrient that fuels cell proliferation. In the former, building on the ideas presented in~\cite{lorenzi2021}, we assume
\beq
 \label{ass:Rred}
R(y,\rho,S) \equiv R(y,\rho), \; R(Y,0) = 0, \; R(0,\rho_M) = 0, \; \partial_\rho R(\cdot,\rho) < 0, \; \partial_y R(y,\cdot) < 0 \; \text{ for } y \in (0,Y),
\eeq 
where $0 < \rho_M <\infty$ is the local carrying capacity of the cell population. In particular, we will focus on the case where
\beq
\label{def:Rred}
R(y,\rho,S) \equiv R(y,\rho) := r(y) - \rho
\eeq
with the function $r(y)$, which models the proliferation rate of cells in the phenotypic state $y$, being such that
$$
r(Y) = 0,  \;\; r(0) = \rho_M, \;\; \partial_y r(y) < 0, \; \text{ for } y \in (0,Y),
$$
so that assumptions~\eqref{ass:Rred} are satisfied. In the latter scenario, we let the function $R(y,\rho,S)$ satisfy the following assumptions
\beq
 \label{ass:R}
R(Y,0,\cdot) = 0, \quad R(0,\rho_m(S),S) = 0, 
\eeq 
\beq
 \label{ass:Rmon}
\partial_\rho R(\cdot,\rho,\cdot) < 0, \;\; \partial_y R(y,\cdot,S) < 0 \; \text{ and } \; \partial_S R(y,\cdot,S) > 0 \; \text{ for } (y,S) \in (0,Y) \times (0,\infty),
\eeq 
where the function $\rho_m(S)$ is such that
\beq
 \label{ass:rhom}
\rho_m(0) = 0, \quad \dfrac{{\rm d} \rho_m(S)}{{\rm d} S} > 0 \; \text{ for } S \in (0,\infty).
\eeq 
In particular, we will focus on the case where
\beq
\label{def:R}
R(y,\rho,S) := r(y,S) - \rho
\eeq
and let the function $r(y,S)$, which models the proliferation rate of cells in the phenotypic state $y$ under the attractant concentration $S$, be such that
$$
r(Y,\cdot) = 0,  \;\; r(0,S) = \rho_m(S), \;\; \partial_y r(y,S) < 0, \;\; \partial_S r(y,S) > 0 \; \text{ for } (y,S) \in (0,Y) \times \mathbb{R}_+,
$$
so that assumptions~\eqref{ass:R}-\eqref{ass:rhom} are satisfied.

The assumptions~\eqref{ass:Rred} and~\eqref{ass:Rmon} on $\partial_\rho R$ correspond to saturating growth, while the assumptions~\eqref{ass:Rred} and~\eqref{ass:Rmon} on $\partial_y R$ model the fact that cells with a higher chemotactic sensitivity are characterised by a lower proliferation rate, for example due to the energetic cost of sensing chemotactic cues~\cite{ni2020} or the acquisition of pro-invasion gene expression and cell signatures being contingent on cell cycle arrest~\cite{lattmann2021}. Furthermore, the assumption~\eqref{ass:Rmon} on $\partial_S R$ translates in mathematical terms to the idea that, when the attractant is considered to be of nutrient-like form, higher levels of attractant availability may correlate with faster cell proliferation. Finally, the function $\rho_m(S)$ models the local carrying capacity of the cell population in the presence of attractant $S$. Assumptions~\eqref{ass:rhom} translate in mathematical terms to the idea that, when the attractant fuels cell proliferation, the higher the concentration of attractant, then the higher the local carrying capacity of the cell population.

\paragraph{Undirected, random cell movement and spontaneous phenotypic changes} The second term on the right-hand side of~\eqref{eq:PDEn} takes into account undirected, random cell movement, which is described through Fick's first law of diffusion with diffusivity $a > 0$, while the third term models the effects of spontaneous, heritable phenotypic changes~\cite{huang2013genetic}, which occur at rate $b >0$.

\paragraph{Environment-induced phenotypic changes} The third term on the left-hand side of~\eqref{eq:PDEn} takes into account the effect of environment-induced, heritable phenotypic changes~\cite{huang2013genetic}. The function $\phi(y,S(x,t))$ models the rate at which cells at position $x$ in the phenotypic state $y$ at time $t$ undergo such changes according to the local environmental conditions, as reflected by the attractant concentration $S(x,t)$. In particular, when the attractant is viewed as a proliferation-fuelling nutrient, it may be natural to suppose that sufficiently high attractant concentrations stimulate phenotypic drift towards a fast-proliferating but minimally-chemotactic state, whereas lower attractant concentrations trigger phenotypic drift towards a slowly-proliferating but highly-chemotactic state. Within this scenario, we let the function $\phi(y,S)$ satisfy the following assumptions
\beq
\label{ass:phi}
\phi(\cdot,S^*) = 0, \quad  \partial_S \phi(y,S^*) \leq 0 \; \text{ for } (y,S) \in (0,Y) \times (0,\infty).
 \eeq 
In assumptions~\eqref{ass:phi}, the parameter $0<S^*<\infty$ is a threshold attractant concentration at which phenotype switching may occur: for $S < S^*$ (for $S > S^*$) a cell may transition into a more-chemotactic (more-proliferative) state, attempting to escape the poor-nutrient region (to exploit nutrient abundance).

\paragraph{Degradation of the attractant} The integral term on the right-hand side of~\eqref{eq:PDES} takes into account the fact that the attractant is degraded by cells in the phenotypic state $y$ at a rate described by the function $\kappa(y,S)$. We consider both a scenario where all cells degrade the attractant at the same rate, independently of their phenotypic state, and a scenario in which cell proliferation may conceivably demand greater consumption of attractant. This latter may be natural when the attractant is considered a nutrient. In the former we assume
\beq
 \label{ass:kappared}
\kappa(y,S) \equiv \kappa(S), \quad \kappa(0) = 0, \quad \dfrac{{\rm d} \kappa(S)}{{\rm d} S}  > 0 \; \text{ for } S \in (0,\infty)
\eeq
and, therefore, the right-hand side of the differential equation~\eqref{eq:PDES} reduces to $\displaystyle{\kappa(S) \, \rho(x,t)}$. On the other hand, in the latter scenario, we let the function $\kappa(y,S)$ satisfy the following assumptions
\beq
 \label{ass:kappa}
\kappa(Y,\cdot) = 0, \;\; \kappa(\cdot,0) = 0, \;\; \partial_S \kappa(y,S) > 0, \;\; \partial_y \kappa(y,S) < 0 \; \text{ for } (y,S) \in (0,Y) \times (0,\infty),
\eeq
which translate in mathematical terms to the idea that cells with a higher proliferation rate consume the attractant at a higher rate. 

\subsection{Object of study}
We focus on a biological scenario in which undirected, random cell movement and spontaneous phenotypic changes occur on a slower time scale compared to chemotactic cell movement and environment-induced phenotypic changes, which in turn occur on a slower time scale compared to cell proliferation and death~\cite{huang2013genetic,smith2004measurement}. To this end, we introduce a small parameter $\e >0$, let 
$$
a := \e^2, \quad b := \e^2, \quad \chi(y) \equiv \e \, \hat \chi(y), \quad  \phi(y,S) \equiv \e \, \hat \phi(y,S)
$$
and then drop the carets from $\hat \chi(y)$ and $\hat \phi(y,S)$.
Moreover, in order to explore the long-time behaviour of the cell population (i.e. the behaviour of the population over many cell generations), we use the time scaling $t \to t / \e$ in~\eqref{eq:PDEn}. Taken together, this gives the following non-local advection-reaction-diffusion equation for the cell population density function $n_{\e}(x,y,t)$ 
\beq
\label{eq:PDEnceps}
 \begin{cases}
\displaystyle{\e \, \partial_t n_{\e} + \e \, \chi(y) \, \partial_x \left(n_{\e}  \, \partial_x S_{\e} \right) + \e \, \partial_y \left(\phi \left(y,S_{\e}\right) n_{\e} \right)= R(y,\rho_{\e},S_{\e}) \, n_{\e} + \e^2 \, \partial^2_{xx} n_{\e} +  \e^2 \, \partial^2_{yy} n_{\e},}
\\\\
\displaystyle{\rho_{\e}(x,t) := \int_{0}^Y n_{\e}(x,y,t) \, {\rm d}y,}
\end{cases}
\eeq
which is coupled with the following differential equation for the concentration of attractant $S_{\e}(x,t)$
\beq
\label{eq:PDESeps}
\displaystyle{\partial_t S_{\e} = - \int_0^Y \kappa(y,S_{\e}) \, n_{\e}(x,y,t) \, {\rm d}y.}
\eeq
Here, $(x,y) \in \mathbb{R} \times (0,Y)$ and~\eqref{eq:PDEnceps} is subject to zero-flux boundary conditions at $y=0$ and $y=Y$.

\newpage
\section{Formal asymptotic analysis}
\label{Sec:Analysis}
In this section, we carry out formal asymptotic analysis of (\ref{eq:PDEnceps}--\ref{eq:PDESeps}).

\subsection{Asymptotic analysis for $\e \to 0$}
Building on the formal method employed in~\cite{lorenzi2021}, which relies on the Hamilton-Jacobi approach developed in~\cite{barles2009concentration,diekmann2005dynamics,lorz2011dirac,perthame2006transport,perthame2008dirac}, we make the real phase WKB ansatz~\cite{barles1989wavefront,evans1989pde,fleming1986pde}
\beq \label{WKB}
n_{\e}(x,y,t) = e^{\frac{u_{\e}(x,y,t)}{\e}},
\eeq
which gives
$$
\partial_x n_{\e} = \frac{\partial_x u_{\e}}{\e} n_{\e}, \quad \partial^2_{xx} n_{\e} = \left(\frac{1}{\e^2} \left(\partial_x u_{\e} \right)^2 + \frac{1}{\e} \partial^2_{xx} u_{\e} \right) n_{\e},
$$
$$
\partial_y n_{\e} = \frac{\partial_y u_{\e}}{\e} n_{\e}, \quad  \partial^2_{yy} n_{\e} = \left(\frac{1}{\e^2} \left(\partial_y u_{\e} \right)^2 + \frac{1}{\e} \partial^2_{yy} u_{\e} \right) n_{\e} \quad \text{and} \quad \partial_t n_{\e} = \frac{\partial_t u_{\e}}{\e} n_{\e}.
$$
Substituting the above expressions into~\eqref{eq:PDEnceps} gives the following Hamilton-Jacobi equation for $u_{\e}(x,y,t)$ 
\begin{eqnarray}
\label{eq:PDEue}
&&\partial_t u_{\e} +  \chi(y) \, \left(\partial_{x} S_{\e} \, \partial_{x} u_{\e} + \e \, \partial^2_{xx} S_{\e}\right) +  \e \, \partial_y \phi \left(y,S_{\e}\right) + \phi \left(y,S_{\e}\right) \partial_y u_{\e} = \nonumber \\
&& \qquad \qquad \qquad \qquad \qquad \qquad \quad R(y,\rho_{\e},S_{\e}) + \left(\partial_x u_{\e} \right)^2 + \e \, \partial^2_{xx} u_{\e} + \left(\partial_y u_{\e} \right)^2 + \e \, \partial^2_{yy} u_{\e}.
\end{eqnarray}
Letting $\e \to 0$ in~\eqref{eq:PDEue} we formally obtain the following Hamilton-Jacobi equation for the leading-order term $u(x,y,t)$ of the asymptotic expansion for $u_{\e}(x,y,t)$
\beq
\label{eq:PDEu}
\partial_t u + \chi(y) \,  \partial_{x} S \, \partial_{x} u + \phi \left(y,S\right) \partial_y u = R(y,\rho,S) + \left(\partial_x u \right)^2 + \left(\partial_y u \right)^2, \quad (x,y) \in \mathbb{R} \times(0,Y).
\eeq
Here, $\rho(x,t)$ is the leading-order term of the asymptotic expansion for $\rho_{\e}(x,t)$, while $S(x,t)$ is the leading-order term of the asymptotic expansion for $S_{\e}(x,t)$. 

\paragraph{Constraint on $u$} When $\rho_{\e} < \infty$ for all $\e > 0$, if $u_{\e}$ is a strictly concave function of $y$ and $u$ is also a strictly concave function of $y$ whose unique maximum point is $\bar{y}(x,t)$, then considering $x \in \mathbb{R}$ such that $\rho(x,t)>0$ (i.e. $x \in \supp(\rho)$) and letting $\e \to 0$ in~\eqref{WKB} formally gives the following constraint on $u$
\beq
\label{eq:ubaryiszero}
u(x,\bar{y}(x,t),t) = \max_{y \in [0,Y]} u(x,y,t) = 0, \quad x \in \supp(\rho),
\eeq
which implies that
\beq
\label{eq:uybaryiszero}
\partial_x u(x,\bar{y}(x,t),t) = 0, \;\; \partial_y u(x,\bar{y}(x,t),t) = 0, \;\; \partial_t u(x,\bar{y}(x,t),t) = 0, \quad x \in \supp(\rho).
\eeq
Note that the system~(\ref{eq:PDEu}-\ref{eq:ubaryiszero}) is a constrained Hamilton-Jacobi equation and $\rho(x,t)>0$ can be regarded as a Lagrange multiplier associated with the constraint~\eqref{eq:ubaryiszero}.

\paragraph{Differential equation for $S$} When $n_\e$ is in the form~\eqref{WKB}, if $u_{\e}$ is a strictly concave function of $y$ and $u$ is also a strictly concave function of $y$ that satisfies the constraint~\eqref{eq:ubaryiszero}, then the following asymptotic result formally holds
$$
\int_{0}^Y \kappa(y,S_{\e}(x,t)) \, n_\e(x,y,t) \, {\rm d}y \xrightarrow[\varepsilon \to 0]{} \kappa(\bar{y}(x,t),S(x,t)) \, \rho(x,t), \quad x \in \mathbb{R}.
$$
Using this asymptotic result along with the differential equation~\eqref{eq:PDESeps}, one finds that $S(x,t)$ formally satisfies the following differential equation
\beq
\label{eq:PDEs}
\partial_t S(x,t) = - \kappa(\bar{y}(x,t),S(x,t)) \, \rho(x,t), \quad x \in \mathbb{R}.
\eeq

\paragraph{Relation between $\bar{y}(x,t)$ and $\rho(x,t)$} Consider $x \in \supp(\rho)$. Evaluating~\eqref{eq:PDEu} at $y=\bar{y}(x,t)$ and using the relations~\eqref{eq:uybaryiszero} we formally find 
\beq
\label{eq:Riszero}
R(\bar{y}(x,t),\rho(x,t),S(x,t)) = 0, \quad x \in \supp(\rho).
\eeq
Under monotonicity assumptions~\eqref{ass:Rred} or~\eqref{ass:Rmon} and~\eqref{ass:rhom}, the relation~\eqref{eq:Riszero} provides, given $S(x,t)$, a one-to-one correspondence between $\bar{y}(x,t)$ and $\rho(x,t)$.

\paragraph{Transport equation for $\bar{y}$} Consider again $x \in \supp(\rho)$. Differentiating~\eqref{eq:PDEu} with respect to $y$, evaluating the resulting equation at $y=\bar{y}(x,t)$ and using the first one and the second one of relations~\eqref{eq:uybaryiszero} yields
\beq
\label{eq:PDEuatbary}
\partial^2_{yt} u(x,\bar{y},t) + \chi(\bar{y}) \,  \partial_{x} S \, \partial^2_{yx} u(x,\bar{y},t) + \phi \left(\bar{y},S\right) \partial^2_{yy} u(x,\bar{y},t)  = \partial_{y} R(\bar{y},S,\rho), \quad x \in \supp(\rho).
\eeq
Moreover, differentiating the second one of relations~\eqref{eq:uybaryiszero} with respect to $t$ and $x$ we find, respectively,
$$
\partial^2_{ty} u(x,\bar{y},t) + \partial^2_{yy} u(x,\bar{y},t) \, \partial_{t} \bar{y}(x,t) = 0 \; \Rightarrow \; \partial^2_{yt} u(x,\bar{y},t) = - \partial^2_{yy} u(x,\bar{y},t) \, \partial_{t} \bar{y}(x,t)
$$
and
$$
\partial^2_{xy} u(x,\bar{y},t) + \partial^2_{yy} u(x,\bar{y},t) \, \partial_{x} \bar{y}(x,t) = 0 \; \Rightarrow \; \partial^2_{yx} u(x,\bar{y},t) = - \partial^2_{yy} u(x,\bar{y},t) \, \partial_{x} \bar{y}(x,t).
$$
Substituting the above expressions of $\partial^2_{yt} u(x,\bar{y},t)$ and $\partial^2_{yx} u(x,\bar{y},t)$ into~\eqref{eq:PDEuatbary}, and using the fact that if $u$ is a strictly concave function of $y$ whose unique maximum point is $\bar{y}(x,t)$ then $\partial^2_{yy} u(x,\bar{y},t) < 0$, gives the following generalised Burgers' equation with source/sink term for $\bar{y}(x,t)$: 
\beq
\label{eq:PDEbary}
\partial_{t} \bar{y} + \chi(\bar{y}) \, \partial_{x} S \, \partial_{x} \bar{y} = \frac{1}{-\partial^2_{yy} u(x,\bar{y},t)} \partial_{y} R(\bar{y},\rho,S) + \phi \left(\bar{y},S\right), \quad x \in \supp(\rho).
\eeq

\begin{remark}
\label{remarkdrift}
If the scaling $\phi(y,S) \equiv \e^2 \, \hat \phi(y,S)$ is considered in the place of the scaling $\phi(y,S) \equiv \e \, \hat \phi(y,S)$ that underlies~\eqref{eq:PDEnceps}, then formal calculations analogous to those above allow one to show that the Hamilton-Jacobi equation~\eqref{eq:PDEu} and the generalised Burgers' equation~\eqref{eq:PDEbary} reduce to their counterparts for the case where $\phi(y,S) \equiv 0$. Hence, formally, under such an alternative scaling, the phenotypic drift will not affect the properties of the solutions to the model equations in the asymptotic regime $\e \to 0$.
\end{remark}

\subsection{Travelling-wave analysis}
In the remainder of this section we will focus on the case where environment-induced phenotypic changes are taken to be negligible, that is, the case where $\phi(y,S) \equiv 0$.
% (i.e. assumption~\eqref{def:phi0} is satisfied). 

\paragraph{Travelling-wave problem} In this case, substituting the travelling-wave ansatz
$$
\rho(x,t) = \rho(z), \quad u(x,y,t) = u(z,y),  \quad \bar{y}(x,t) = \bar{y}(z) \quad \text{and} \quad S(x,t) = S(z),
$$
with $z = x - c \, t$ and $c > 0$, into~\eqref{eq:PDEu}-\eqref{eq:uybaryiszero}, \eqref{eq:Riszero}, \eqref{eq:PDEbary} and \eqref{eq:PDEs} gives 
\beq
\label{eq:TWu}
\left(\chi(y) \, S' - c \right) \partial_z u  = R(y,\rho,S) + (\partial_z u)^2 + (\partial_y u)^2, \quad (z,y) \in \mathbb{R} \times (0,Y),
\eeq
\beq
\label{eq:ubaryiszeroTW}
u(z,\bar{y}(z)) = \max_{y \in [0,Y]} u(z,y) = 0, \quad \partial_z u(z,\bar{y}(z)) = 0, \quad \partial_y u(z,\bar{y}(z)) = 0, \quad z \in \supp{\left(\rho \right)}, 
\eeq
\beq
\label{eq:TWRiszero}
R(\bar{y}(z),\rho(z),S(z)) = 0, \quad z \in \supp{\left(\rho \right)},
\eeq
\beq
\label{eq:TWbary}
\left(c - \chi(\bar{y}) S' \right) \bar{y}'  = \frac{1}{\partial^2_{yy} u(z,\bar{y})} \partial_{y} R(\bar{y},\rho,S), \quad z \in \supp{\left(\rho \right)}
\eeq
and
\beq
\label{eq:TWs}
c \, S' = \kappa(\bar{y},S) \, \rho, \quad z \in \mathbb{R}.
\eeq
Focussing on a biological scenario in which the concentration of the attractant is at the equilibrium value $0<S_0 < \infty$ prior to cell invasion, we require the following asymptotic condition to be satisfied
\beq
\label{eq:TWBCS}
\lim_{z \to + \infty}  S(z) = S_0.
\eeq 
Moreover, building on the results presented in~\cite{lorenzi2021}, we seek monotonically increasing solutions of the differential equation~\eqref{eq:TWbary} subject to the asymptotic condition 
\beq
\label{eq:TWBCy}
\lim_{z \to - \infty}  \bar{y}(z) =0.
\eeq

\paragraph{Preliminary observations} Under assumptions~\eqref{ass:kappared} or~\eqref{ass:kappa}, since $\rho(z)$ is non-negative, the solutions to the problem (\ref{eq:TWs}-\ref{eq:TWBCS}) satisfy the following properties
\beq
\label{eq:TWSinc}
0 \leq S(z) \leq S_0, \quad S'(z) \geq 0, \quad z \in \mathbb{R}.
\eeq
We also note that, since $\partial_\rho R < 0$ (\emph{cf.} assumptions~\eqref{ass:Rred} or~\eqref{ass:Rmon}), differentiating the relation~\eqref{eq:TWRiszero} with respect to $z$ gives the following differential relation
\beq
\label{eq:TWRziszero}
\rho' = - \dfrac{1}{\partial_{\rho} R(\bar{y},\rho,S)} \Big(\partial_y R(\bar{y},\rho,S) \, \bar{y}'  + \partial_{S} R(\bar{y},\rho,S) \, S' \Big), \quad z \in \supp{\left(\rho \right)}.
\eeq

\subsubsection{Travelling-wave solutions under assumptions~\eqref{ass:Rred}}
\label{subsec:TW1}
We start by noting that when cell proliferation is independent from the concentration of the attractant, i.e. when assumptions~\eqref{ass:Rred} hold, the asymptotic condition~\eqref{eq:TWBCy} along with the relation~\eqref{eq:TWRiszero} gives
\beq
\label{eq:TWBCrho}
\lim_{z \to - \infty} \rho(z) = \rho_M.
\eeq

\paragraph{Minimal wave speed} If assumptions~\eqref{ass:Rred} hold then $\partial_y R <0$. Hence, since $\partial^2_{yy} u(z,\bar{y})<0$, the differential equation~\eqref{eq:TWbary} along with the monotonicity property~\eqref{eq:TWSinc} allows one to conclude that the following condition needs to hold for $\bar{y}(z)$ to be a monotonically increasing function:
\beq
\label{minc}
c > \sup_{z \in \supp{\left(\rho \right)}} \chi(\bar{y}(z)) \, S'(z) =: c_{{\rm min}}.
\eeq

\paragraph{Shape of travelling-wave solutions and position of the leading edge} 
If assumptions~\eqref{ass:Rred} are satisfied then $\partial_y R <0$, $\partial_{\rho} R <0$ and $\partial_{S} R \equiv 0$. Hence, the differential relation~\eqref{eq:TWRziszero} yields
\beq
\label{sgnbaryrho}
\sgn\left(\rho'(z)\right) = -\sgn\left(\bar{y}'(z)\right), \quad z \in \supp{\left(\rho \right)}.
\eeq
In conclusion, if $c$ meets condition~\eqref{minc} then 
\beq
\label{eq:TWbaryincrhodec}
\bar{y}'(z) > 0 \quad \text{and} \quad \rho'(z) < 0, \quad z \in \supp{\left(\rho \right)}.
\eeq

\sskip
Moreover, if assumptions~\eqref{ass:Rred} hold then $R(Y,0)=0$. Hence, the relation~\eqref{eq:TWRiszero}, the monotonicity results~\eqref{eq:TWbaryincrhodec} and the asymptotic relation~\eqref{eq:TWBCrho} allow one to conclude that the position of the leading edge of a travelling-wave solution $\bar{y}(z)$ that satisfies the differential equation~\eqref{eq:TWbary} subject to the asymptotic condition~\eqref{eq:TWBCy} coincides with the unique point $\ell \in \mathbb{R}$ such that $\bar{y}(\ell)=Y$ and 
\beq
\label{propsupprho1}
\rho(z) =0  \; \text{for } z \in (\ell, \infty).
\eeq

\sskip
Finally, since $S(z)$ satisfies (\ref{eq:TWs}-\ref{eq:TWBCS}), the following properties hold
\beq
\label{eq:TWBCS1}
S(z) = S_0  \; \text{for } z \in (\ell, \infty), \quad S'(z) > 0 \; \text{for } z \in (-\infty, \ell), \quad \lim_{z \to - \infty} S(z) = 0.
\eeq  
In the case where the function $\kappa(y,S)\equiv\kappa(S)$ satisfies assumptions~\eqref{ass:kappared} and is bounded for $S \in [0,S_0]$, properties~\eqref{eq:TWBCS1} are obtained by studying the behaviour of the solutions to (\ref{eq:TWs}-\ref{eq:TWBCS}) using the property~\eqref{propsupprho1} and the monotonicity property~\eqref{eq:TWbaryincrhodec} of $\rho(z)$ along with the asymptotic relation~\eqref{eq:TWBCrho}. On the other hand, in the case where the function $\kappa(y,S)$ satisfies assumptions~\eqref{ass:kappa} and is bounded for $(y, S) \in [0,Y] \times [0,S_0]$, properties~\eqref{eq:TWBCS1} are obtained by studying the behaviour of the solutions to (\ref{eq:TWs}-\ref{eq:TWBCS}) using the monotonicity property~\eqref{eq:TWbaryincrhodec} of $\bar{y}(z)$ along with the asymptotic condition~\eqref{eq:TWBCy} -- which ensures that $\bar{y}(z) < Y$ on $\supp(\rho)$ and, therefore, $\kappa(\bar{y},S)>0$ on $\supp(\rho) \cap \supp(S)$ -- and both the property~\eqref{propsupprho1} and the monotonicity property~\eqref{eq:TWbaryincrhodec} of $\rho(z)$ along with the asymptotic relation~\eqref{eq:TWBCrho}.  

\begin{remark}
When assumptions~\eqref{ass:Rred} hold and the function $R(y,\rho,S)$ is defined via~\eqref{def:Rred}, the relation~\eqref{eq:TWRiszero} gives 
\beq
\label{eq:TWRiszerored1}
\rho(z)=r(\bar{y}(z)), \quad z \in \supp(\rho).
\eeq
Substituting the relation~\eqref{eq:TWRiszerored1} into the differential equation~\eqref{eq:TWs} yields
$$
c \, S'(z) = \kappa(\bar{y}(z), S(z)) \, r(\bar{y}(z)), \quad z \in \supp(\rho)
$$
and thus condition~\eqref{minc} can be rewritten as 
\beq
\label{eq:minws1}
c > \sup_{z \in \supp{\left(\rho \right)}} \sqrt{\chi(\bar{y}(z)) \, \kappa(\bar{y}(z), S(z)) \,  r(\bar{y}(z))} =: c_{{\rm min}}.
\eeq
\end{remark}

\subsubsection{Travelling-wave solutions under assumptions~\eqref{ass:R}-\eqref{ass:rhom}}
\label{subsec:TW2}
We start by noting that when the attractant is viewed as a nutrient fuelling cell proliferation, i.e. when assumptions~\eqref{ass:R}-\eqref{ass:rhom} hold, the relation~\eqref{eq:TWRiszero} implies that 
\beq
\label{eq:supportofrhoandS}
\supp{\left(\rho \right)} \subseteq \supp{\left(S\right)}.
\eeq

\paragraph{Minimal wave speed} If assumptions~\eqref{ass:R}-\eqref{ass:rhom} hold then $\partial_y R(y,\cdot,S) <0$ for all $S \in (0,S_0]$. Hence, arguments analogous to those used in Section~\ref{subsec:TW1} alongside properties~\eqref{eq:TWSinc} and~\eqref{eq:supportofrhoandS} allow one to conclude that $c$ needs to satisfy condition~\eqref{minc} for $\bar{y}(z)$ to be a monotonically increasing function.

\paragraph{Shape of travelling-wave solutions and position of the leading edge} Under assumptions~\eqref{ass:R}-\eqref{ass:rhom}, if $c$ meets condition~\eqref{minc} then the differential equation~\eqref{eq:TWbary} along with the property~\eqref{eq:supportofrhoandS} yields 
\beq
\label{eq:TWbaryincr2}
\bar{y}'(z) > 0, \quad z \in \supp{\left(\rho \right)}.
\eeq

\sskip
Moreover, if assumptions~\eqref{ass:R}-\eqref{ass:rhom} hold then $R(Y,0,\cdot)=0$. Hence, the relation~\eqref{eq:TWRiszero}, the monotonicity property~\eqref{eq:TWSinc} along with the asymptotic condition~\eqref{eq:TWBCS}, the property~\eqref{eq:supportofrhoandS} and the monotonicity result~\eqref{eq:TWbaryincr2} allow one to conclude that the position of the leading edge of a travelling-wave solution $\bar{y}(z)$ that satisfies the differential equation~\eqref{eq:TWbary} subject to the asymptotic condition~\eqref{eq:TWBCy} coincides with the unique point $\ell \in \mathbb{R}$ such that $\bar{y}(\ell)=Y$ and
\beq
\label{propsupprho2}
\rho(z) =0  \; \text{for } z \in (\ell, \infty).
\eeq

\sskip
Finally, since $S(z)$ satisfies (\ref{eq:TWs}-\ref{eq:TWBCS}), properties~\eqref{eq:TWBCS1} hold and
\beq
\label{eq:TWBCSrho2}
\lim_{z \to -\infty} \rho(z) = 0.
\eeq
In the case where the function $\kappa(y,S)\equiv\kappa(S)$ satisfies assumptions~\eqref{ass:kappared} and is bounded for $S \in [0,S_0]$, properties~\eqref{eq:TWBCS1} and~\eqref{eq:TWBCSrho2} are obtained by studying the behaviour of the solutions to (\ref{eq:TWs}-\ref{eq:TWBCS}) using properties~\eqref{eq:supportofrhoandS} and~\eqref{propsupprho2}. On the other hand, in the case where the function $\kappa(y,S)$ satisfies assumptions~\eqref{ass:kappa} and is bounded for $(y, S) \in [0,Y] \times [0,S_0]$, properties~\eqref{eq:TWBCS1} and~\eqref{eq:TWBCSrho2} are obtained by studying the behaviour of the solutions to (\ref{eq:TWs}-\ref{eq:TWBCS}) using the monotonicity result~\eqref{eq:TWbaryincr2} along with the asymptotic condition~\eqref{eq:TWBCy} -- which ensures that $\bar{y}(z) < Y$ on $\supp(\rho)$ and, therefore, $\kappa(\bar{y},S)>0$ on $\supp(\rho) \cap \supp(S)$ -- and both properties~\eqref{eq:supportofrhoandS} and~\eqref{propsupprho2}.

\begin{remark}
\label{rem:critpoint}
The results obtained so far under assumptions~\eqref{ass:R}-\eqref{ass:rhom} ensure that
$$
\partial_y R(\bar{y}(z),\rho(z),S(z)) \, \bar{y}'(z) < 0 \quad \text{and} \quad \partial_{S} R(\bar{y}(z),\rho(z),S(z)) \, S'(z) > 0, \quad z \in \supp{\left(\rho \right)}.
$$
These facts along with the differential relation~\eqref{eq:TWRziszero}, the property~\eqref{propsupprho2} and the asymptotic property~\eqref{eq:TWBCSrho2} support the idea that the total cell density $\rho(z)$ will have one single non-degenerate critical point, which will be a maximum point.
\end{remark}

\begin{remark}
When assumptions~\eqref{ass:R}-\eqref{ass:rhom} hold and the function $R(y,\rho,S)$ is defined via~\eqref{def:R}, the relation~\eqref{eq:TWRiszero} gives 
\beq
\label{eq:TWRiszerored2}
\rho(z)=r(\bar{y}(z),S(z)), \quad z \in \supp(\rho).
\eeq
Substituting the relation~\eqref{eq:TWRiszerored2} into the differential equation~\eqref{eq:TWs} yields
$$
c \, S'(z) = \kappa(\bar{y}(z),S(z)) \, r(\bar{y}(z), S(z)), \quad z \in \supp(\rho)
$$
and thus condition~\eqref{minc} can be rewritten as 
\beq
\label{eq:minws2}
c > \sup_{z \in \supp{\left(\rho \right)}} \sqrt{\chi(\bar{y}(z)) \, \kappa(\bar{y}(z),S(z)) \, r(\bar{y}(z),S(z))} =: c_{{\rm min}}.
\eeq
\end{remark}

\newpage
\section{Numerical simulations}
\label{Sec:Numerics}
\paragraph{Set-up of the numerical simulations} We numerically investigate the dynamics of (\ref{eq:PDEnceps}--\ref{eq:PDESeps}). Note that for the numerical method utilised here it is necessary to restrict the physical domain to the closed interval $[0, L]$. At the boundaries we impose lossless conditions on $n_{\e}(x,y,t)$, i.e. we also set zero-flux boundary conditions across the physical boundaries at $x = 0$ and $x=L$ in addition to those previously stated for the phenotype boundaries at $y = 0$ and $y=Y$. We choose $Y=1$ and, unless stated otherwise, $L=20$. 

\sskip
Initially, a relatively small population of cells is localised along the $x=0$ boundary, uniformly distributed across the phenotype space. Attractant is initially set at a constant positive level. The initial conditions could therefore represent, as an example, a population of bacteria cells clustered at one end of a pipette, microchannel or a tube that contains some attractant or nutrient. Specifically, we implement the following set of initial conditions
\[
n_{\e} (0,x,y) \equiv n_{\e}(0,x) := N_0 \exp (- \zeta x)\,, \quad S_{\e} (0,x) \equiv S_0\,,
\]
where $0<S_0 < \infty$ models the equilibrium value of the concentration of attractant prior to cell invasion. We set $N_0=0.1, S_0=1$ and $\zeta = 30$ throughout the simulations. 

\paragraph{Numerical method} The numerical scheme invokes a Method of Lines approach: we discretise in both physical and phenotype space, specifying a uniform mesh of spacing $\Delta x$ and $\Delta y$, respectively, and integrate the resulting high-dimensional ordinary differential equation (ODE) system in time. Spatial and phenotype movement terms are discretised in conservative term, with central differencing applied to diffusive terms and first-order upwind scheme for advective terms (an alternative higher-order upwind scheme provided no significant improvement of accuracy when balanced against cost efficiency). The scheme was encoded in MATLAB and {\tt{ode45}} was the default choice for integrating the ODE system, with absolute and relative error tolerances both set at $10^{-8}$. We set $\Delta x = 0.005$ and $\Delta y = 0.01$ (corresponding to 4000 and 100 grid points for the physical and phenotype domains, respectively, when $L=20$ and $Y=1$). Numerical controls included both decreasing and increasing $\Delta x$ and $\Delta y$ values by factors of two, setting lower error tolerances and employing alternative time-stepping schemes to solve the ODE system. We note that the general approach above is based on standard diffusion-taxis schemes (e.g. \cite{hundsdorfer2003}), but extended to include phenotypic variation. 

\begin{figure}[t!]
\begin{center}
\includegraphics[width=\textwidth]{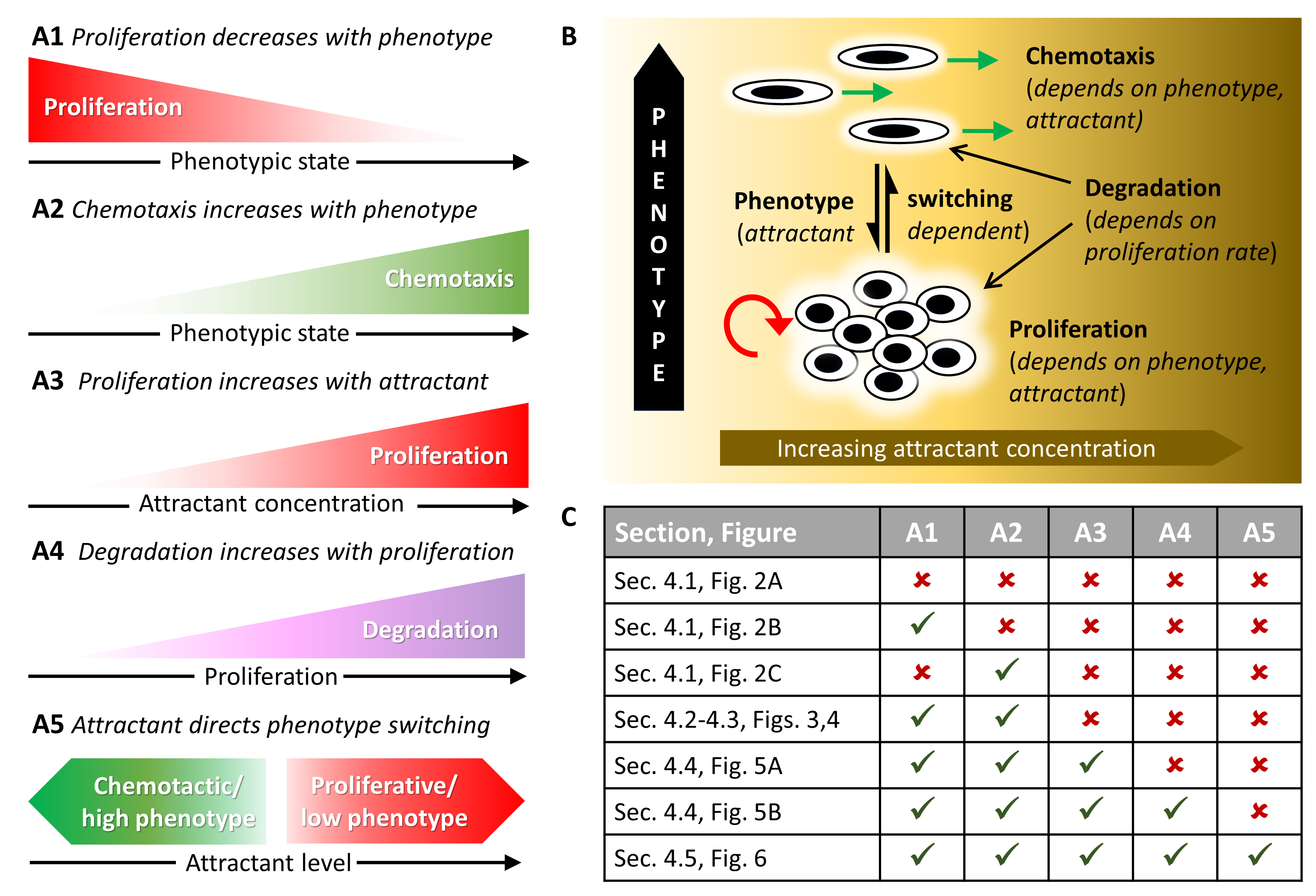}
\end{center}
\caption{{\bf Model scenarios considered in the numerical investigation}. {\bf A} 
Functional relationships considered for the various model terms: {\bf A1} Proliferation rate decreases with the phenotypic state, $y$; {\bf A2} Chemotactic sensitivity increases with the phenotypic state, $y$; {\bf A3} Proliferation rate increases with the concentration of the attractant, $S$; {\bf A4} Rate of attractant degradation/consumption by cells increases with the proliferation rate of cells; {\bf A5} Attractant concentration can trigger phenotype transitioning, with low (high) attractant concentrations resulting in a transition to phenotypic states represented by higher (lower) values of $y$. 
% higher (lower) phenotype states
{\bf B} Figure indicating the key behaviours built into the chemotaxis/proliferation trade-off model, where phenotypic variants range from minimally-chemotactic but highly-proliferative cells (red circular arrows) to minimally-proliferative but highly-chemotactic cells (green straight arrows) that migrate up the attractant gradient (background colour). Cells degrade the attractant (fading edging) and can transition through phenotypic states according to the concentration level. {\bf C} Table demonstrating how the various functional relationships illustrated in {\bf A} are steadily included into the model framework, with reference to the appropriate figure and section.}\label{fig:figure1}
\end{figure}

\newpage
\paragraph{Numerical exploration}
Our numerical exploration investigates dynamics under a number of functional dependencies,
conforming to the general assumptions set out in Section \ref{Sec:Model}. In particular, we consider the following list of relationships, see also Figure~\ref{fig:figure1}A:
\begin{enumerate}
%\item[(A1)] cells in lower phenotype state are more proliferative, Figure~\ref{fig:figure1}A1;
\item[(A1)] cells in phenotypic states represented by lower values of $y$ are more proliferative, Figure~\ref{fig:figure1}A1;
%\item[(A2)] cells in a higher phenotype state are more chemotactic, Figure~\ref{fig:figure1}A2;
\item[(A2)] cells in phenotypic states represented by higher values of $y$ are more chemotactic, Figure~\ref{fig:figure1}A2;
\item[(A3)] the proliferative rate of cells increases with the level of the attractant, e.g. the attractant is considered a nutrient that fuels cell proliferation, Figure~\ref{fig:figure1}A3;
\item[(A4)] the rate of attractant degradation/consumption increases with the rate of proliferation, Figure~\ref{fig:figure1}A4;
%\item[(A5)] the concentration of attractant determines the direction of phenotype switching towards low or high phenotype states, Figure~\ref{fig:figure1}A5.
\item[(A5)] the concentration of attractant determines the direction of phenotype switching towards phenotypic states represented by lower or higher values of $y$, Figure~\ref{fig:figure1}A5.
\end{enumerate}
The simultaneous application of (A1) and (A2) is referred to as the chemotaxis-proliferation trade-off model and a schematic showing the possible features of the model is provided in Figure~\ref{fig:figure1}B. The table in Figure~\ref{fig:figure1}C indicates which of the above set of assumptions are included within various simulation suites, with reference to the appropriate figure and section.

\subsection{Base scenarios}
Prior to the principal study, we explore dynamics in certain simple scenarios. We neglect the effect of environment-induced phenotypic changes, i.e. we assume
\begin{equation} \label{def:phi0}
\phi(y,S) \equiv 0,
\end{equation}
and consider the cell phenotypic state 
\begin{enumerate}
\item[{\bf A}] has no impact on proliferation, chemotaxis and attractant degradation, i.e.
\beq
\label{def:cA}
\chi(y) \equiv \alpha, \quad R(y,\rho,S) \equiv R(\rho) := \beta - \rho, \quad \kappa(y,S) \equiv \kappa(S) := \gamma \, S,
\eeq
\end{enumerate}
or 
\begin{enumerate}
\item[{\bf B}] impacts only on proliferation, i.e. 
\beq
\label{def:cB}
\chi(y) \equiv \alpha, \quad R(y,\rho,S) \equiv R(y,\rho) := \beta \, (1-y) - \rho, \quad \kappa(y,S) \equiv \kappa(S) := \gamma \, S,
\eeq
\end{enumerate}
or 
\begin{enumerate}
\item[{\bf C}] impacts only on chemotaxis, i.e.
\beq
\label{def:cC}
\chi(y) \equiv \alpha y, \quad R(y,\rho,S) \equiv R(\rho) := \beta - \rho, \quad \kappa(y,S) \equiv \kappa(S) := \gamma \, S.
\eeq
\end{enumerate}
Here, $\alpha>0$, $\beta>0$ and $\gamma>0$. Scenario {\bf A} corresponds to the case of a phenotypically-homogeneous population.

\begin{figure}[t!]
\begin{center}
\includegraphics[width=\textwidth]{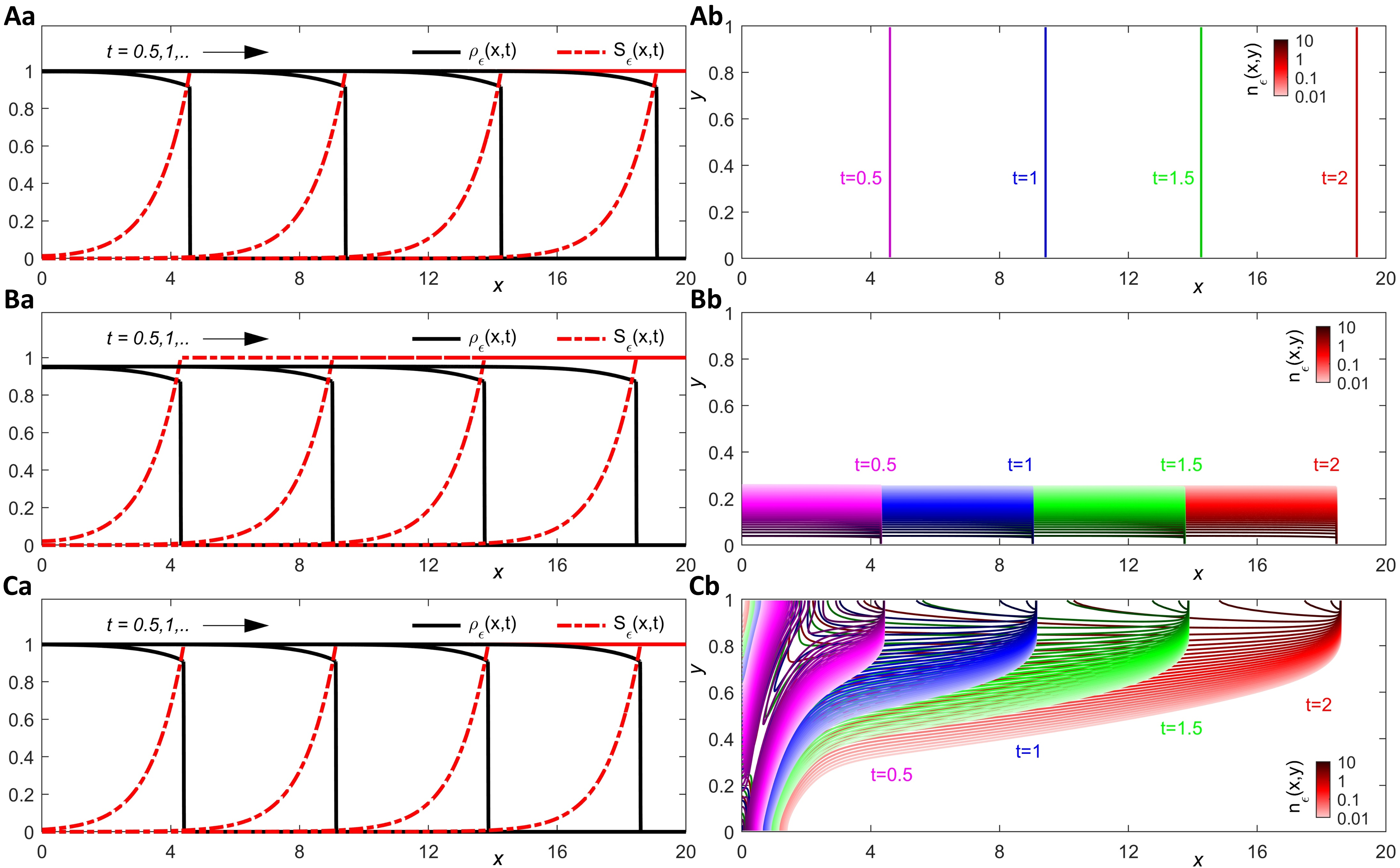}
\end{center}
\caption{{\bf Evolution to travelling-wave solutions in base scenarios.} Model functions are defined via~\eqref{def:phi0} and {\bf A}~\eqref{def:cA}, {\bf B}~\eqref{def:cB} or {\bf C}~\eqref{def:cC}. {\bf a} Total cell density $\rho_{\e}(x,t)$ (solid black lines) and attractant concentration $S_{\e}(x,t)$ (dash-dot red lines) plotted at time increments of 0.5, $t=0.5,1,1.5,2$. {\bf b} Cell population density $n_{\e}(x,y,t)$, where light to dark contour lines indicate increasing density and different tones discriminate the distributions at (magenta) $t=0.5$, (blue) $t=1$, (green) $t=1.5$, (red) $t=2$. Parameters are set $\e = 0.01$, $\alpha=10$, $\beta=1$ and $\gamma = 10$.}\label{fig:figure2}
\end{figure}

\sskip
Typical dynamics are displayed in Figure~\ref{fig:figure2}, where we plot in {\bf a} the total cell density, $\rho_\e(x,t)$, and attractant concentration, $S_\e(x,t)$, and in {\bf b} the cell population density, $n_\e(x,y,t)$, at progressive times. With respect to the total cell density and attractant concentration, solutions under all three cases {\bf A}-{\bf C} display very similar dynamics, where we observe evolution to travelling-wave profiles of almost identical shapes and speeds. Notably, though, there is significant variation when it comes to the phenotype distribution of cells. When the cell phenotypic state neither impacts on proliferation nor chemotaxis, Figure~\ref{fig:figure2}{\bf A}, the model reduces to a simple one-dimensional model for a homogeneous cell population and we note a uniform distribution of cells across phenotype space. When the cell phenotypic state only impacts on proliferation (i.e. all cells have the same chemotactic sensitivity), we observe dominance by those phenotypic variants with the most rapid proliferation, Figure~\ref{fig:figure2}{\bf B}, a result of their obvious competitive advantage. This is flipped when the cell phenotypic state impacts only on chemotactic sensitivity, Figure~\ref{fig:figure2}{\bf C}, where the highly-chemotactic cells now gain a competitive advantage through advancing to the leading edge and dominating proliferation in that region.

\subsection{Chemotaxis/proliferation trade-off}
Having established dynamics in the aforementioned base cases, we explore dynamics under the {\em chemotaxis/proliferation trade-off} (i.e. the fact that, succinctly, proliferative cells are less chemotactic and vice-versa). Note that at this first stage we assume that the attractant does not impact on cellular proliferation and that all cells degrade it at the same rate. Moreover, we neglect the effect of environment-induced phenotypic changes. Hence, we let the model functions satisfy assumptions~\eqref{ass:chi}, \eqref{ass:Rred}, \eqref{ass:kappared} and~\eqref{def:phi0}. We will focus on the case where $R(y,\rho,S)$ is defined via~\eqref{def:Rred} and use the following (linear) definitions to carry out numerical simulations
\begin{equation} \label{eq:linear}
\chi(y) := \alpha y \, , \quad r(y) := \beta \left(1-y\right)\,, \quad \kappa(y,S) \equiv \kappa(S) := \gamma S\,.  
\end{equation}
Here, $\alpha>0$ denotes the maximum chemotactic sensitivity (i.e. the chemotactic sensitivity of cells in the phenotypic state $y=Y$), $\beta>0$ indicates the maximum proliferation rate (i.e. the proliferation rate of cells in the phenotypic state $y=0$), and $\gamma>0$ describes the rate at which a cell degrades the attractant. Note that when $r(y)$ is defined via~\eqref{eq:linear}, the local carrying capacity of the cell population is $\rho_M = \beta$.

\begin{figure}[t!]
\begin{center}
\includegraphics[width=\textwidth]{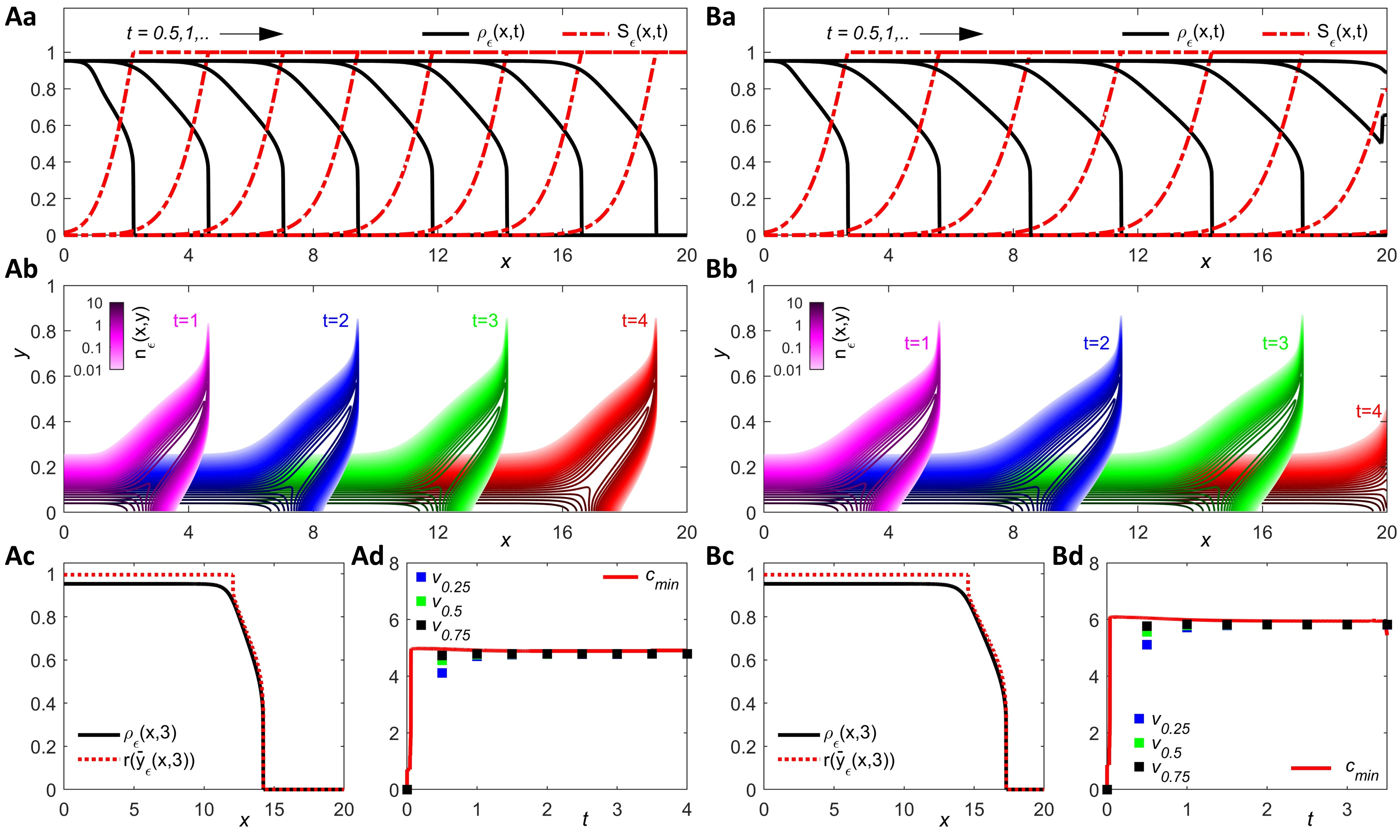}
\end{center}
\caption{{\bf Evolution to travelling-wave solutions under chemotaxis/proliferation trade-off.} Model functions are defined via~\eqref{def:phi0}, \eqref{def:Rred} and~\eqref{eq:linear} with {\bf A} $\alpha = 10$ or {\bf B} $\alpha = 15$. {\bf a} Total cell density $\rho_{\e}(x,t)$ (solid black lines) and attractant concentration $S_{\e}(x,t)$ (dashed red lines) plotted at times $t=0,0.5,1,\hdots,4$. {\bf b} Cell population density $n_{\e}(x,y,t)$, where light to dark contour lines indicate increasing density and different tones discriminate the distributions at (magenta) $t=1$, (blue) $t=2$, (green) $t=3$, (red) $t=4$. {\bf c} Comparison between $\rho_{\e}(x,3)$ (solid black) and $r(\bar{y}_{{\e}}(x,3))$ (dotted red). {\bf d} Comparison between front propagation speeds $v_{S_i}(t)$ computed numerically via~\eqref{eq:wfprspS} for three thresholds $S_i$ and the minimal wave speed $c_{{\rm min}}$ computed numerically via the formula~\eqref{eq:minws1}. Note that in {\bf Bd} the time is truncated at $t=3.5$, that is, the time at which the leading edge reaches the boundary of the physical domain at $x=20$. Other parameters are set $\e=0.01$, $\beta=1$ and $\gamma = 10$.}\label{fig:figure3}
\end{figure}

\sskip
Typical dynamics are displayed in Figure~\ref{fig:figure3}, where we note that all parameters have been fixed with the exception of the maximum chemotactic sensitivity $\alpha$, which we set at (Figure~\ref{fig:figure3}{\bf A}) $\alpha = 10$ or (Figure~\ref{fig:figure3}{\bf B}) $\alpha = 15$. We plot in {\bf a} the total cell density, $\rho_{\e}$, and attractant concentration, $S_{\e}$, and in {\bf b} the cell population density, $n_{\e}$, at progressive times. In {\bf c}-{\bf d} the simulation results are compared with the results of a formal asymptotic analysis of (\ref{eq:PDEnceps}--\ref{eq:PDESeps}) for $\e \to 0$, presented in Section~\ref{Sec:Analysis}.

\sskip
Numerical solutions again support the formation of travelling waves, whereby a population of cells expands from source and degrades the attractant, Figures~\ref{fig:figure3}{\bf Aa,Ba}. Compared to the simpler base scenarios, the incorporation of a chemotaxis/proliferation trade-off lead cells to be non-uniformly distributed across both physical and phenotype space, Figures~\ref{fig:figure3}{\bf Ab,Bb}. More precisely, we observe a relatively small subpopulation of highly-chemotactic but minimally-proliferative cells (i.e. cells in phenotypic states $y\approx Y$) that becomes concentrated towards the front of the invading wave, while rapidly-proliferating but minimally-chemotactic cells (i.e. cells in phenotypic states $y\approx 0$) make up the bulk of the population in the rear. Such structuring of the population is reminiscent of the ``leader-follower'' type behaviour described in a number of instances of collective cell migration, e.g.~\cite{haeger2015,mayor2016,mercedes2021}. Intuitively, efficient gradient-following by the most-chemotactic cells leads to their positioning at the front of the wave, where the attractant is highest. This competitive advantage is temporary, however, with the greater proliferation of less-chemotactic cells allowing those cells to eventually dominate the rear. Increasing the maximum chemotactic sensitivity $\alpha$ -- compare Figure~\ref{fig:figure3}{\bf A} with Figure~\ref{fig:figure3}{\bf B} -- extends the width of the region of the invading wave in which highly-chemotactic cells are found, and there is a corresponding increase in the invasion speed. 

\sskip
As a more precise numerical test for possible travelling-wave dynamics, we track the propagation speeds $v_{S_i}(t)$ of the wavefront $S_\e$ for various threshold levels $S_i$, where
\beq
\label{eq:wfprspS}
v_{S_i}(t) := \dfrac{x_{S_i}(t)}{t} \quad \text{with} \quad x_{S_i}(t) \quad \text{such that} \quad S_{\e}(x_{S_i}(t),t) = S_i\,.
\eeq
Notably, we observe evolution towards a common and constant speed, suggesting that solutions indeed converge to a form with constant speed and shape, corroborating our supposition of travelling waves, Figures~\ref{fig:figure3}{\bf Ad,Bd}. We remark that an equivalent tracking of propagation speeds of the wavefront $\rho_\e$ for various threshold levels $\rho_i$ yields quasi-identical results.

\sskip
The plots in Figure~\ref{fig:figure3} also indicate that, when $\e$ is sufficiently small and after a transient interval, the population density function $n_{\e}(x,y,t)$ becomes concentrated as a sharp Gaussian with maximum at a point $\bar{y}_{\e}(x,t)$, which corresponds to the dominant phenotype, for all $x \in \supp(\rho_{\e})$, Figures~\ref{fig:figure3}{\bf Ab,Bb}. In agreement with the results of the formal analysis carried out in Section~\ref{Sec:Analysis}: 
\begin{itemize}
\item the maximum point $\bar{y}_{\e}(x,t)$ behaves like a compactly supported and monotonically increasing travelling front that connects $y=0$ to $y=Y$;
\item the total cell density $\rho_{\e}(x,t)$ behaves like a one-sided compactly supported and monotonically decreasing travelling front that connects $\rho_M$ to $0$; 
\item the attractant concentration $S_{\e}(x,t)$ behaves like a travelling front that increases monotonically on $\supp(\rho_{\e})$ and connects $0$ to $S_0$, Figures~\ref{fig:figure3}{\bf Aa,Ba} and Figures~\ref{fig:figure3}{\bf Ab,Bb}. 
\end{itemize}
This means that the formal results concerning the shape of travelling-wave solutions and the position of the leading edge presented in Section~\ref{subsec:TW1} hold. Moreover, again after a transient interval, we find a good quantitative agreement between $\rho_{\e}(x,t)$ and $r(\bar{y}_{{\e}}(x,t))$ for all $x \in \supp(\rho_{\e})$, which means that the relation~\eqref{eq:TWRiszerored1} holds as well, Figures~\ref{fig:figure3}{\bf Ac,Bc}. Finally, the value of the wave speed is in agreement with the value of the minimal wave speed $c_{{\rm min}}$ which is computed numerically via the formula~\eqref{eq:minws1}, Figures~\ref{fig:figure3}{\bf Ad,Bd}. We verified that, {\it ceteris paribus}, the smaller is the value of $\e$, then the better such a quantitative agreement between numerical and analytical results. 
\begin{remark}
Travelling-wave dynamics appear to extend to other parameter regimes of the model, such as moving away from the small $\e$ assumption implicit in the formulation (\ref{eq:PDEnceps}--\ref{eq:PDESeps}) ({\it cf.} Figures~\ref{fig:figureappendix}{\bf A} in Appendix~\ref{appendix:supfig}) or when chemotaxis is taken to be negligible ({\it cf.} Figure~\ref{fig:figureappendix}{\bf B} in Appendix~\ref{appendix:supfig}). However, in such regimes we can no longer expect close correspondence with the results of the formal analysis ({\it cf.} Figures~\ref{fig:figureappendix}{\bf Ac,Bc} and Figures~\ref{fig:figureappendix}{\bf Ad,Bd} in Appendix~\ref{appendix:supfig}).
\end{remark}

\sskip
Summarising, these initial simulations demonstrate that the acquisition of a highly-chemotactic phenotype acts to accelerate invasion through positioning cells that express this phenotype at the leading edge of the wave. These ``exploratory'' cells are subsequently replaced by highly-proliferative cells in the rear.

\subsection{Different forms of chemotaxis/proliferation balance}
We extend the study of chemotaxis/proliferation trade-off carried out in the previous section by investigating the impact of {\em chemotaxis/proliferation balance} (i.e. different forms of balance between phenotype-dependent chemotaxis and phenotype-dependent proliferation) on the invasion profile. In analogy with the previous section we define the function $\phi(y,S)$ via~\eqref{def:phi0}, the function $\kappa(y,S)$ via~\eqref{eq:linear} and the function $R(y,\rho,S)$ via~\eqref{def:Rred}, but choose the following nonlinear forms for the functions $\chi(y)$ and $r(y)$: 
\begin{equation} \label{eq:nonlinear}
\chi(y) := \eta \left(1+\theta y\right)^p \quad \mbox{and} \quad r(y) := \frac{\beta}{\left(1+\theta y\right)^q}, \quad p,q\ge 0.
\end{equation}
Here, $\eta>0$ now denotes the minimal chemotactic sensitivity (i.e. the chemotactic sensitivity of cells in the phenotypic state $y=0$) and $\beta>0$ indicates again the maximum proliferation rate (i.e. the proliferation rate of cells in the phenotypic state $y=0$). Note that in this case the assumption~\eqref{ass:Rred} on $R(Y,0)$ does not hold, since $r(Y)>0$. Note also that selecting $p = q = 0$ (or $\theta = 0$) eliminates any variation with phenotype and the model is reduced to a one-dimensional model for a homogeneous cell population (see Figure \ref{fig:figure2}{\bf A} for typical dynamics). We define the product $\chi(y)r(y)$ as the {\em combined proliferative-chemotactic potential} of cells in the phenotypic state $y$. In scenarios where $p=q$, the combined proliferative-chemotactic potential will be constant as the phenotypic state $y$ varies. We refer to such scenarios as evenly balanced.  
\begin{figure}[t!]
\begin{center}
\includegraphics[width=\textwidth]{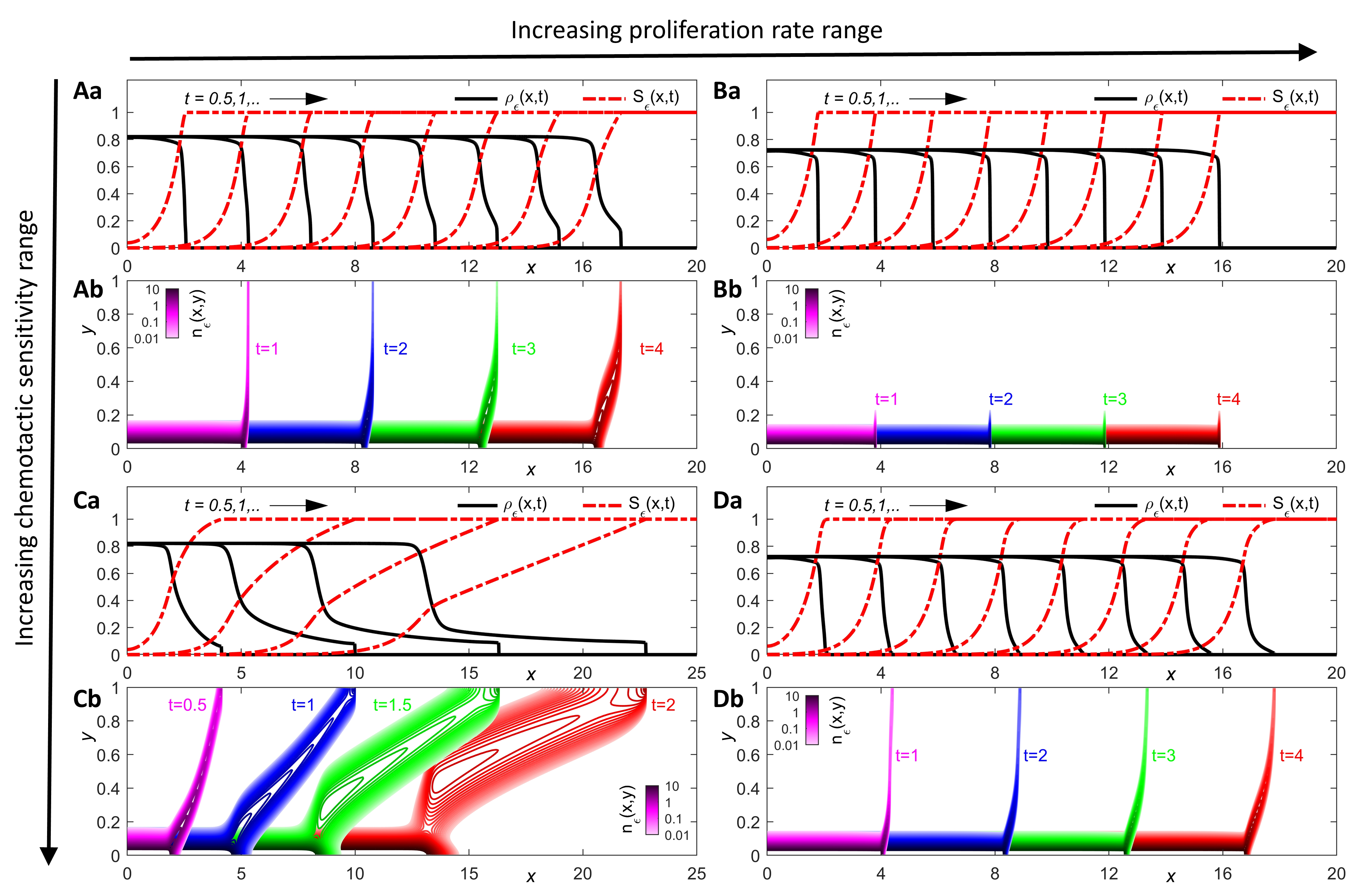}
\end{center}
\caption{{\bf Evolution to travelling-wave solutions under different forms of chemotaxis/proliferation balance.} Model formulation and parameter values as in Figure~\ref{fig:figure3}, but with functions $\chi(y)$ and $r(y)$ defined via~\eqref{eq:nonlinear} with {\bf A} $(p,q) = (1,1)$, {\bf B} $(p,q) = (1,2)$, {\bf C} $(p,q) = (2,1)$ or {\bf D} $(p,q) = (2,2)$. {\bf a} Total cell density $\rho_{\e}(x,t)$ (solid black lines) and attractant concentration $S_{\e}(x,t)$ (dash-dot red lines) plotted at time increments of 0.5, $t=0.5,1,\hdots, 4$ (except {\bf C}, where the simulation is stopped at $t=2$). {\bf b} Cell population density $n_{\e}(x,y,t)$, where light to dark contour lines indicate increasing density and different tones discriminate the distributions at the different times indicated. Note that in {\bf C} the physical domain is extended to $L = 25$ in order to show the phenomenon of ``wave stretching'' that occurs under parameter combination {\bf C}. Other parameters are set $\eta=2$ and $\theta = 9$.}\label{fig:figure4}
\end{figure}

\sskip
The plots in Figures~\ref{fig:figure4}{\bf A-D} summarise the numerical results obtained for $\beta = 1$, $\eta = 2$ and $\theta = 9$, and four different $(p,q)$ combinations: {\bf A} $(p,q)=(1,1)$, {\bf B} $(p,q)=(1,2)$, {\bf C} $(p,q)=(2,1)$, and {\bf D} $(p,q) = (2,2)$. The choice $\theta=9$ ensures that, as $y$ increases from 0 to $Y$, with $Y=1$, the chemotactic sensitivity increases by a factor of $10^p$ and the proliferation rate decreases by a factor of $10^q$. Parameter combinations {\bf A} and {\bf D} both correspond to evenly balanced scenarios, i.e. both the chemotactic sensitivity and proliferation rate vary over the same order of magnitude (factors of 10 in {\bf A} and 100 in {\bf D}). The dynamics of these two cases appear somewhat similar. We observe the formation of fronts in which a subpopulation of highly-chemotactic cells (i.e. those in phenotypic states $y \approx Y$) leads at the invasive front, while fast-proliferating cells (i.e. those in phenotypic states $y \approx 0$) are found in the rear. Furthermore, we observe a comparable size and span of the highly-chemotactic subpopulation and similar overall invasion rates, Figures~\ref{fig:figure4}{\bf A},{\bf D}.

\sskip
Parameter combination {\bf B} tilts the balance away from even, so that while proliferation decreases by a factor of 100, chemotactic sensitivity only increases by a factor of 10. Overall, the combined proliferative-chemotactic potential monotonically decreases with $y$. A dramatic reduction in the size of the highly-chemotactic subpopulation is observed and invasion is reduced, Figure~\ref{fig:figure4}{\bf B}. Here, the more-chemotactic cells are unable to break sufficiently free at the front: their comparatively poor proliferative capacity leads to these cells being quickly overcome by the highly-proliferative subpopulation encroaching from the rear of the wave.

\sskip
Parameter combination {\bf C} also tilts the balance away from even, but in the opposite direction: while the proliferation rate decreases by a factor of 10, chemotactic sensitivity now increases by a factor of 100 and the combined proliferative-chemotactic potential is monotonically increasing with $y$. This reverse shift leads to a phenomenon of ``wave stretching'', Figure~\ref{fig:figure4}{\bf C}. Specifically, we observe a lower density plateau of exploratory cells that rapidly stretches outwards at the wave front. Here, highly-chemotactic cells break free from the mass and their higher proliferative potential (with respect to equivalent subpopulations in Figures~\ref{fig:figure4}{\bf A,B,D}) increases the time before these cells are replaced by highly-proliferative phenotypic variants. 

\sskip 
Overall, the results in this section reveal how the balance between chemotaxis and proliferation is crucial to the size and structure of the exploratory subpopulation at the leading front of invading waves. 

\subsection{Attractant-dependent proliferation}
We move to scenarios in which {\em attractant-dependent proliferation} occurs (i.e. the attractant is viewed in the light of a nutrient that fuels cell proliferation). Hence, we suppose the fitness function $R(y,\rho_{\e},S_{\e})$ satisfies assumptions~\eqref{ass:R}-\eqref{ass:rhom}. We will focus on the case where $R(y,\rho,S)$ is defined via~\eqref{def:Rred} and use the following (linear) definition to carry out numerical simulations
\beq
\label{def:r}
r(y,S) := \beta S \left(1-y\right).
\eeq
Here, $\beta>0$ denotes again the maximum proliferation rate (i.e. the proliferation rate of cells in the phenotypic state $y=0$). We again define the chemotactic sensitivity function $\chi(y)$ via~\eqref{eq:linear}, and consider both the case where all cells degrade the attractant at an equal rate (i.e. the function $\kappa(y,S_{\e})$ satisfies assumptions~\eqref{ass:kappared}) and the case of {\em attractant-dependent proliferation with linked degradation}, that is, the case where cell proliferation demands greater consumption of the attractant/nutrient (i.e. the function $\kappa(y,S_{\e})$ satisfies assumptions~\eqref{ass:kappa}). In the former we again define the function $\kappa$ via~\eqref{eq:linear}, while in the latter case we let attractant degradation be proportional to the cell proliferation rate (i.e. $\kappa(y,S_{\e}) \propto r(y,S_\e)$) and thus use the following definition 
\beq
\label{def:k}
\kappa(y,S) := \gamma S \left(1-y\right),
\eeq
where $\gamma>0$ indicates again the maximum rate at which a cell degrades the attractant (i.e. the rate at which the attractant is degraded by cells in the phenotypic state $y=0$).

\begin{figure}[t!]
\begin{center}
\includegraphics[width=\textwidth]{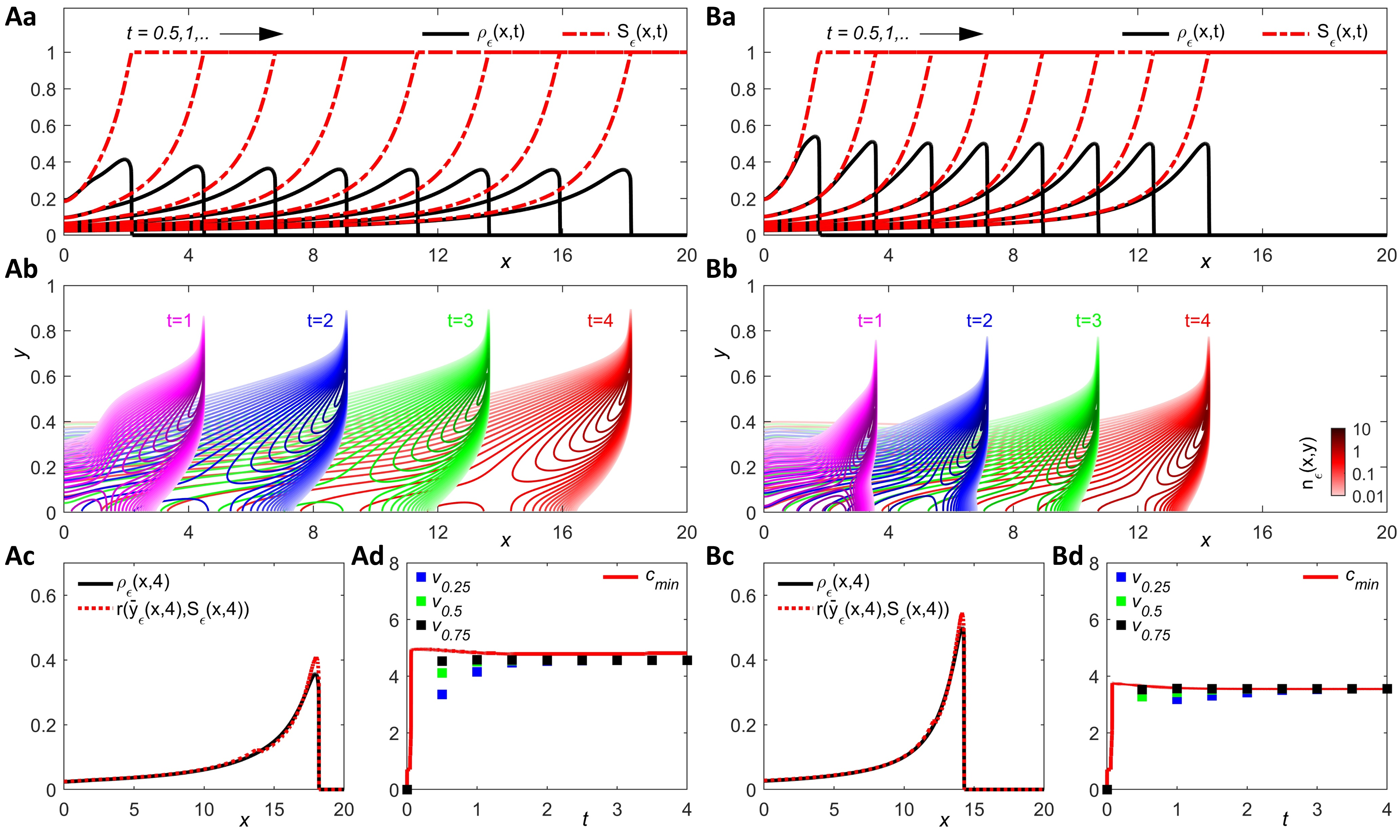}
\end{center}
\caption{{\bf Evolution to travelling-wave solutions under attractant-dependent proliferation.} Functions $\phi(y,S)$ and $\chi(y,S)$ are defined via~\eqref{def:phi0} and~\eqref{eq:linear}, the function $R(y,\rho,S)$ is defined via~\eqref{def:R} and~\eqref{def:r}, and the function $\kappa(y,S)$ is defined via {\bf A}~\eqref{eq:linear} or {\bf B}~\eqref{def:k}. {\bf a} Total cell density $\rho_{\e}(x,t)$ (solid black lines) and attractant concentration $S_{\e}(x,t)$ (dashed red lines) plotted at times $t=0,0.5,1,\hdots,4$. {\bf b} Cell population density $n_{\e}(x,y,t)$, where light to dark contour lines indicate increasing density and different tones discriminate the distributions at (magenta) $t=1$, (blue) $t=2$, (green) $t=3$, (red) $t=4$. {\bf c} Comparison between $\rho_{\e}(x,4)$ (solid black) and $r(\bar{y}_{{\e}}(x,4), S_{{\e}}(x,4))$ (dotted red). {\bf d} Comparison between front propagation speeds $v_{S_i}$ computed numerically via~\eqref{eq:wfprspS} for three thresholds $S_i$ and the minimal wave speed $c_{{\rm min}}$ computed numerically via the formula~\eqref{eq:minws2}. Parameters are set $\e = 0.01$, $\alpha=10$, $\beta=1$ and $\gamma=10$.}\label{fig:figure5}
\end{figure}

\sskip
Typical dynamics are displayed in Figure~\ref{fig:figure5}, where in {\bf A} all cells degrade the attractant at an equal rate, and in {\bf B} attractant degradation is proportional to the cell proliferation rate. As in Figure~\ref{fig:figure3}, we plot {\bf a} the total cell density, $\rho_{\e}$, and attractant concentration, $S_{\e}$, and {\bf b} the cell population density, $n_{\e}$, at progressive times. In {\bf c}-{\bf d} the simulation results are again compared with the results of a formal asymptotic analysis of (\ref{eq:PDEnceps}--\ref{eq:PDESeps}) for $\e \to 0$, presented in Section~\ref{Sec:Analysis}.

\sskip
Simulations once again support the formation of travelling waves. The total cell density $\rho_{\e}$ now takes the form of a travelling pulse whereby cells are concentrated at the forefront of the wave, Figures~\ref{fig:figure5}{\bf Aa,Ba}. We remark that simulations performed over a longer timeframe imply that both $\rho_\e$ and $S_\e$ at the back of the wave converge to zero (data not shown). As before, we find a ``leader-follower'' type structuring in which a subpopulation of cells with high chemotactic sensitivity (i.e. cells in phenotypic states $y \approx Y$) is concentrated at the leading edge, while fast-proliferating cells with low chemotactic sensitivity (i.e. cells in phenotypic states $y \approx 0$) are found in the rear subpopulation, Figures~\ref{fig:figure5}{\bf Ab,Bb}. 

\sskip
Depletion of the attractant (or nutrient) diminishes the proliferation of all cells, and hence the total cell density in the rear is steadily reduced. A comparison between the plots in Figure~\ref{fig:figure5}{\bf Aa,Ad} and the plots in Figure~\ref{fig:figure5}{\bf Ba,Bd} supports the idea that, in the case where attractant degradation is proportional to the cell proliferation rate, there is a more concentrated travelling pulse and reduced invasion speed. This reduced invasion can be attributed to the fact that the most-chemotactic cells are no longer degrading the attractant, leading in turn to a shallower attractant gradient and reduced invasion at the leading edge. Consequently, the more-proliferative cells have more time to advance towards the leading edge, where the attractant concentration is higher and faster growth can occur. In turn the highly-chemotactic subpopulation is diminished in comparison to the case in which all cells degrade the attractant at an equal rate, as cells in this subpopulation become more rapidly out-competed.

\sskip
Similarly to the plots in~Figure~\ref{fig:figure3}, the plots in Figure~\ref{fig:figure5} also indicate that, when $\e$ is sufficiently small and after a transient interval, the population density function $n_{\e}(x,y,t)$ becomes concentrated as a sharp Gaussian with maximum at a point $\bar{y}_{\e}(x,t)$, which corresponds to the dominant phenotype, for all $x \in \supp(\rho_{\e})$, Figures~\ref{fig:figure5}{\bf Ab,Bb}. In agreement with the results of the formal analysis carried out in Section~\ref{Sec:Analysis}, 
\begin{enumerate}
\item the maximum point $\bar{y}_{\e}(x,t)$ behaves like a compactly supported and monotonically increasing travelling front that connects $y=0$ to $y=Y$, 
\item the total cell density $\rho_{\e}(x,t)$ behaves like a one-sided compactly supported travelling pulse with one single non-degenerate critical point, which is a maximum point, and
\item the attractant concentration $S_{\e}(x,t)$ behaves like a travelling front that increases monotonically on $\supp(\rho_{\e})$ and connects $0$ to $S_0$, Figures~\ref{fig:figure5}{\bf Aa,Ba} and Figures~\ref{fig:figure5}{\bf Ab,Bb}. 
\end{enumerate}
Once again this indicates that the formal results concerning the shape of travelling-wave solutions and the position of the leading edge presented in Section~\ref{subsec:TW2} hold. Moreover, again after a transient interval, we find a a good quantitative agreement between $\rho_{\e}(x,t)$ and $r(\bar{y}_{{\e}}(x,t), S_{\e}(x,t))$ for all $x \in \supp(\rho_{\e})$, which means that the relation~\eqref{eq:TWRiszerored2} holds as well, Figures~\ref{fig:figure5}{\bf Ac,Bc}. Finally, the value of the wave speed is in agreement with the value of the minimal wave speed $c_{{\rm min}}$ which is computed numerically via the formula~\eqref{eq:minws2}, Figures~\ref{fig:figure5}{\bf Ad,Bd}. We verified that, {\it ceteris paribus}, the smaller is the value of $\e$, then the better such a quantitative agreement between numerical and analytical results. 

\sskip
Overall, the results of this section demonstrate that inclusion of attractant-dependent proliferation leads to travelling-pulse type spatial dynamics of the cells, where the pulse is composed of a subpopulation of highly-chemotactic explorer-type cells at the front, and more-proliferative cells in the rear.

\subsection{Attractant-dependent phenotypic drift}
In the scenarios above, phenotypic drift has been taken as negligible: we restricted the assumption on the function $\phi$ to that specified in~\eqref{def:phi0}. In this section we extend our study to incorporate phenotype drift. There are a number of range of choices for such an investigation, and here we confine ourselves to a simple extension of the attractant-dependent proliferation scenario considered in the previous section. Specifically, viewing the attractant in the light of a nutrient that fuels cell proliferation, it could be natural to suppose an adaptive-type response in which cells that find themselves in a poor-nutrient (i.e. low-attractant) region acquire a more-chemotactic phenotype. Highly-chemotactic cells that have moved into a good-nutrient (i.e. high-attractant) region, on the other hand, are assumed to transition into a more-proliferative phenotypic state. 

\sskip
Hence, we suppose the fitness function $R(y,\rho_{\e},S_{\e})$ satisfies assumptions~\eqref{ass:R}-\eqref{ass:rhom}, again taking the form~\eqref{def:R} with $r(y,S_{\e})$ defined via~\eqref{def:r}. Furthermore, we again consider the linear chemotactic sensitivity function $\chi(y)$ defined via~\eqref{eq:linear}, and consider the case where degradation of the attractant is proportional to the cell proliferation rate, i.e. $\kappa(y,S)$ is given by~\eqref{def:k}. In other words, the formulation will be identical to that used for the numerical simulations in Figure~\ref{fig:figure5}{\bf B}, but extended to include {\em attractant-dependent phenotypic drift}. In particular, we let the function $\phi(y,S_\e)$ satisfy the assumptions~\eqref{ass:phi} and, for simplicity, we will set
\beq
\label{def:phi}
\phi(y,S) \equiv \phi(S) := \varphi \, (S^*-S).
\eeq 
As previously described, $0<S^*<S_0$ represents a threshold attractant concentration at which phenotype switching occurs. Moreover, $\varphi>0$ defines a simple rate parameter that modulates the rate at which cells are able to shift their phenotype.

\begin{figure}[t!]
\begin{center}
\includegraphics[width=\textwidth]{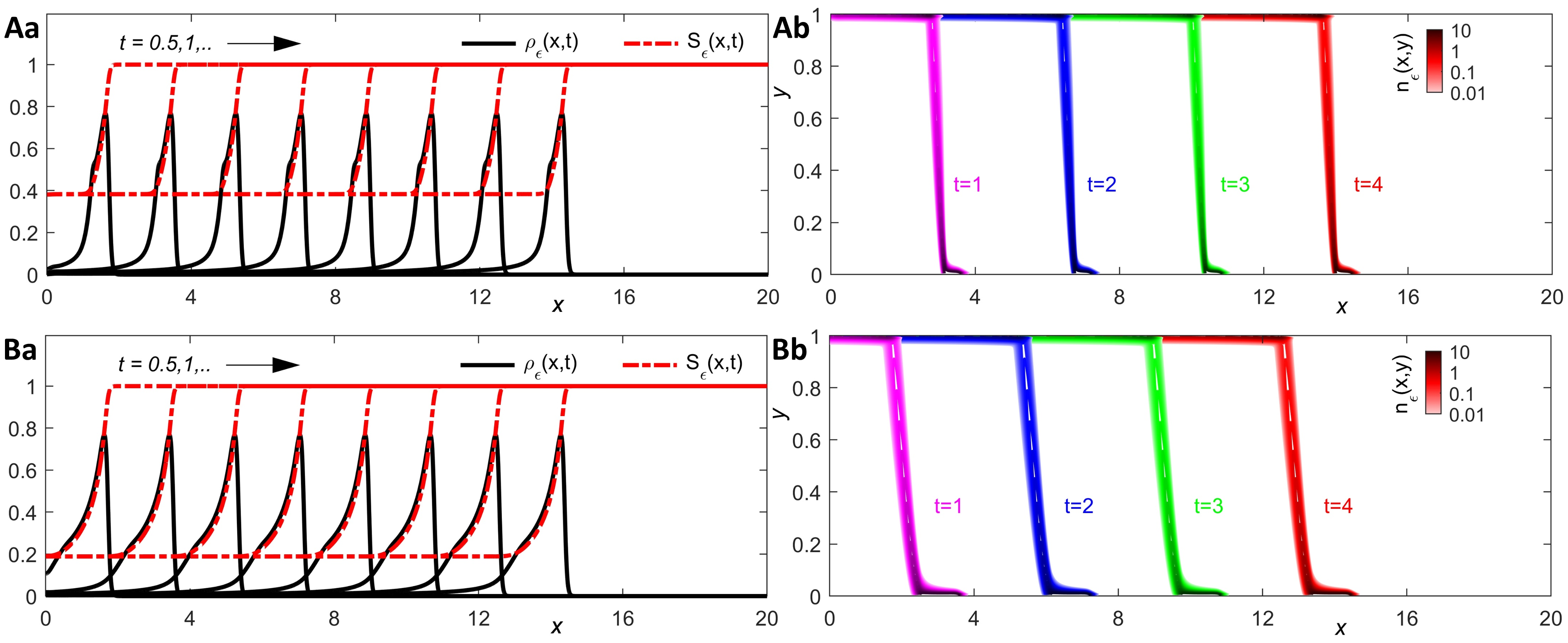}
\end{center}
\caption{{\bf Evolution to travelling-wave solutions under attractant-dependent phenotypic drift.} Model formulation and parameter values as in Figure~\ref{fig:figure5}{\bf B}, but with the function $\phi(y,S)$ defined via~\eqref{def:phi} with {\bf A} $(\varphi,S^*) = (1,0.5)$ or {\bf B} $(\varphi,S^*) = (1,0.25)$. {\bf a} Total cell density $\rho_{\e}(x,t)$ (solid black lines) and attractant concentration $S_{\e}(x,t)$ (dashed red lines) plotted at times $t=0,0.5,1,\hdots,4$. {\bf b} Cell population density $n_{\e}(x,y,t)$, where light to dark contour lines indicate increasing density and different tones discriminate the distributions at (magenta) $t=1$, (blue) $t=2$, (green) $t=3$, (red) $t=4$.}\label{fig:figure6}
\end{figure}

\sskip
Results from a set of representative cases are displayed in Figure \ref{fig:figure6}, and as a point of direct comparison we refer to Figure~\ref{fig:figure5}{\bf B} (corresponding to $\varphi = 0$).  Dynamics are shown for parameter choices {\bf A} $(S^*,\varphi)=(1,0.5)$ and {\bf B} $(S^*,\varphi)=(1,0.25)$. Overall, we continue to observe travelling-wave behaviour, with the cell density arranged into travelling-pulse form. There are, however, a number of distinctions when compared to the dynamics displayed in Figure~\ref{fig:figure5}{\bf B}. 

\sskip
First, we observe a significantly more concentrated pulse, see Figures~\ref{fig:figure6}{\bf Aa,Ba}. This consolidation is driven by phenotypic drift: in the rear of the wave, the attractant becomes depleted and at a certain point drops below the threshold level $S^*$ (i.e. $S_{\e}(x,t) < S^*$). This triggers the phenotype transition, with previously-proliferating cells becoming chemotactic and migrating up the attractant gradient in the direction of the invasive front. Eventually they cross back into a high-attractant region (i.e. $S_{\e}(x,t) > S^*$) and the reverse transition takes place, transitioning back into a highly-proliferative phenotypic state. 

\sskip
Second, we note a reversal in the cell phenotype distribution across the wavefront. Earlier invasion waves were driven by highly-chemotactic cells at the front and fast-proliferating cells building up in the rear. Environment-dependent phenotypic drift switches this, with the highly-chemotactic cells in the rear (escaping the low-attractant regime) and fast-proliferating cells at the front (exploiting the attractant abundance). A significant reduction in $\varphi$ (e.g. setting $\varphi = \e$ with $\e=0.01$, results not shown) can slow the advective drift to the point that transitions are not fast enough to effect this switch, and the travelling-wave profile returns to one resembling the profile in Figure~\ref{fig:figure5}{\bf B}. This is to be expected on the basis of the formal asymptotic results presented in Section~\ref{Sec:Analysis} (see Remark~\ref{remarkdrift}).

\sskip
Finally, we note that the attractant concentration at the rear of the wave no longer converges to zero, but rather to some positive value. In the rear of the wave, the conversion of cells into the chemotactic phenotype and their subsequent escape from the low-attractant region drives the cell density behind the wave to negligible levels. The absence of cells reduces attractant degradation to negligible levels, and the attractant level instead settles to some value smaller than $S^*$. Setting a lower threshold level $S^*$, naturally, lowers the point at which the phenotypic switch occurs and the attractant behind the wave is reduced to a lower level -- compare Figure~\ref{fig:figure6}{\bf A} with Figure~\ref{fig:figure6}{\bf B}. 

\section{Discussion and research perspectives}
\label{Sec:Discussion}
We have extended a Patlak-Keller-Segel type model for chemotaxis-invasion processes to incorporate phenotypic heterogeneity of the population. The phenotype enters as a continuous variable that directly impacts on certain characteristics, where here we have concentrated on a {\em chemotaxis/proliferation trade-off} scenario: a spectrum of behaviours from more-proliferative and less-chemotactic to more-chemotactic and less-proliferative. Across a variety formulations -- for example, regarding the attractant as a nutrient or allowing the attractant to dictate the direction of phenotype transition -- we observe the generation of travelling waves in which phenotypes are structured across the support of the wave. When phenotypic transitions arise through random fluctuations, structuring is principally determined by the competition between invasive and proliferative processes: highly-chemotactic cells locally dominate at the invasive front, but are eventually overwhelmed by more-proliferative cells that encroach from the rear. Including phenotypic drift, though, can alter this structuring. In particular, we found that an attractant-dependent phenotypic drift led to a reversal in which more-chemotactic cells are found at the rear, and more-proliferative cells are located at the front. 

\sskip
We note that the current study has eschewed a specific biological context, with the assumptions sufficiently generic for the migration/invasion of micro-organisms, during development, wound healing, cancer {\em etc}; our ``cells'' could also be reinterpreted as a species in an ecological context. While this maximises generality, it does limit us to making more general conclusions. Future investigations may therefore benefit from a focussed biological application, thereby refining the model assumptions and allowing certain key questions to be addressed. 

\sskip
A natural such application would be to explore the extent to which phenotypic variation benefits a microbial population in a changing nutrient landscape, an area of significant current interest~\cite{gasperotti2020}. Recent studies have explored energy investments of {\em E. coli} bacteria, which display a negative correlation between chemotactic gene promoters and population growth rate~\cite{ni2020}, in line with the assumption of the chemotaxis/proliferation trade-off considered here. Notably, exposure to a poor-nutrient environment was found to lead to an increase in investment in motility~\cite{ni2020}, reminiscent of our incorporation of a drift towards more-chemotactic phenotypic states below a critical nutrient level. In other relevant studies, subjecting {\em E. coli} populations to microfluidic ``T-mazes'' highlights how populations are phenotypically-structured with respect to their chemotactic sensitivity, such that those with stronger sensitivity are capable of more deeply infiltrating the maze structure~\cite{salek2019}. Parametrising and fitting the model to such set-ups would facilitate an investigation into how phenotypic changes alter a microbial population's robustness to fluctuating-nutrient environments.

\sskip
In the context of cancer invasion, the ``go-or-grow'' hypothesis  posits a dichotomy between proliferation and migration and was conceived following observations of glioma cell behaviour~\cite{giese1996}. Experimental tests into its wider applicability remain ambiguous, with data both supportive (e.g.~\cite{giese1996,hoek2008}) and against (e.g.~\cite{corcoran2003,vittadello2020}). Nevertheless, substantial interest remains and numerous mathematical models have incorporated its central tenet (e.g.~\cite{hatzikirou2012,pham2012,stepien2018,zhigun2018}), typically through supposing two cell-state variables and incorporating switching between states. On a similar note, a recent study has investigated how malignant invasion varies in nutrient-depleted environments, via two cell-state models where cells can have distinct chemotactic sensitivity and/or nutrient-dependent growth~\cite{kimmel2020}. The chemotaxis/proliferation trade-off explored here is, fundamentally, of ``go-or-grow'' nature but our overall framework extends to a broader and continuous spectrum of phenotypic states across a population, as well as allowing environment-dependent phenotype transitions.

\sskip
Neural crest migration offers a natural application within developmental biology, and more widely provides a paradigm system for studying cell invasion. One integrated experimental-theoretical approach has focussed on chick cranial neural crest cell migration, suggesting a process in which cells follow a vascular endothelial growth factor (VEGF) attractant gradient self-generated through their uptake of VEGF~\cite{giniunaite2020}. An agent-based modelling approach suggested that distinct cell phenotypes (termed ``leaders'' and ``followers'') were required for successful migration, differing (amongst other factors) in their response to the gradient of VEGF~\cite{giniunaite2020}. While these agent-based models have been extended to include a continuous spectrum of phenotypes~\cite{schumacher2019}, the modelling approach here (appropriately modified) could allow a fully continuous approach to be adopted. More widely, the classification of cells into follower- or leader-types has been widely adopted in collective cell migration processes, although it has been noted that using such terminologies based purely on position (e.g. leaders at the front, followers at the back) could obscure the considerable variation and subtlety through which different populations generate coordinated migration~\cite{theveneau2017}. In this context, we note that the single addition of attractant-dependent phenotypic drift was sufficient to dramatically alter the phenotypic distribution across the invading population. 

\sskip
Turning to more general extensions, a number of further studies could yield deeper insight into how the dynamical interplay between spatial and evolutionary mechanisms shape biological invasion processes. First, our analysis has focussed on a very specific scaling of the model~(\ref{eq:PDEn}-\ref{eq:PDES}), leading to the rescaled model~(\ref{eq:PDEnceps}-\ref{eq:PDESeps}). Our primary motivation, here, was to facilitate the formal asymptotic approach adopted in Section~\ref{Sec:Analysis}, and indeed we observed agreement between the results of our numerical simulations and the analytical predictions, provided model parameters conformed with the scaling assumptions. While the validity of the analysis itself weakens when shifting away from this scaling regime, at a broader level simulations indicate similar overall behaviour: solutions typically evolve to travelling-wave solutions (e.g. see Figures~\ref{fig:figureappendix}{\bf Aa-c} in Appendix~\ref{appendix:supfig}). Extending the study to other parameter regimes would clarify the degree to which travelling waves are universally expected, or whether structurally different dynamics can be obtained (e.g. accelerating fronts~\cite{berestycki2015existence,bouin2012invasion,bouin2017super,lorenzi2021}). Another natural and relatively straightforward generalisation would be to consider diffusion and decay of the attractant. The exclusion of these processes here was primarily motivated by a desire for simplicity, and we would naturally expect that certain key features would carry over -- indeed, results from simulations of~(\ref{eq:PDEnceps}--\ref{eq:PDESeps}) modified to include attractant diffusion are shown in Figures~\ref{fig:figureappendix}{\bf Ca-c} in Appendix~\ref{appendix:supfig} -- when these additional elements of biological complexity are incorporated into the model. 

\sskip
Regarding possible applications to mechanobiology, a further track to follow would be to include the effects of cell-cell mechanical interactions, mechanical interactions between cells and components of the extracellular matrix (i.e. the network of extracellular macromolecules that provide cells with structural support and segregate tissues) and haptotaxis (i.e. directional cell movement in response to adhesive components in the extracellular matrix)~\cite{arduino2015multiphase,chauviere2010model,ciarletta2013mechanobiology,giverso2018mechanical,giverso2018nucleus,preziosi2009multiphase,roux2016prediction,verdier2009rheological}. Investigations here could disentangle the role of chemotactic movement, phenotypic adaptation and mechanical interactions at the cellular scale in the growth and remodelling of living tissues. From a mathematical point of view, this would require further development of the numerical and formal asymptotic methods employed here so that a similar mathematical study into the spatial eco-evolutionary dynamics of cells could be conducted. 

\sskip
Intriguing questions arise if extending the modelling to explore pattern formation scenarios. Patlak-Keller-Segel models are well known for their self-organising capacity, first explored in the context of {\em Dictyostelium discoideum} self organisation~\cite{keller1970} and subsequently proposed across a broad gamut of applications from microbiology to social sciences (see~\cite{painter2019} for a review). The classical chemotaxis-driven instability results from a coupling between chemotaxis and population production of the attractant, thereby generating a positive feedback that brings a dispersed population into a cluster. Under phenotypic variation, a population may have a range of chemotactic sensitivities and attractant production rates, with intriguing consequences on the qualitative and quantitative features of patterning. For example, from the pattern initiation scenario, do certain phenotypic variations act to limit or promote pattern formation? From a mathematical perspective, when does global existence or finite-time blow up occur? Extension of the model in this direction could provide some interesting insights.

\sskip
As noted above, (numerical) travelling fronts and/or travelling pulses have been found across a broad range of scenarios, though it is worth noting that all formulations have included (phenotypic dependent) population growth. Population growth may be minimal or absent over the timescale of invasion, and a mathematical question arises as to whether travelling waves will form in such scenarios. The formation of a sustained pulse in the absence of growth demands that each population member keeps pace with the wave, but whether this is possible is uncertain: cells at different positions will be exposed to a different environment, so cells at the rear exposed to shallower gradients risk losing contact, and the pulse may disperse. Travelling waves in models relying on the formulation of Keller and Segel~\cite{keller1971a} (see also the review~\cite{wang2013}) circumvent this through a ``logarithmic'' sensitivity, but this effectively gives a cell a capacity to detect and respond to infinitesimally shallow gradients, which may be  biochemically infeasible (see the discussion in~\cite{xue2011}). It would be interesting to see whether certain phenotype trade-offs could lead to travelling waves in the formulation~(\ref{eq:PDEn}-\ref{eq:PDES}) of the model, when the net proliferation rate $R(y,\rho,S)$ is assumed to be negligible. 

\sskip
Taken together, the results of this initial study point the way towards novel compelling research directions for the mathematical modelling of chemotaxis-driven invasions.

\bigskip
{\bf Acknowledgements:} T.L. gratefully acknowledges support from the MIUR grant ``Dipartimenti di Eccellenza 2018-2022''. K.J.P. acknowledges ``MIUR–Dipartimento di Eccellenza'' funding to the Dipartimento Interateneo di Scienze, Progetto e Politiche del Territorio (DIST).

\bibliography{references}

\newpage

\appendix

\renewcommand{\thesection}{A.\arabic{section}} 
\section{Supplementary figures}
\label{appendix:supfig}

 \beginsupplement

\begin{figure}[h!]
\begin{center}
\includegraphics[width=\textwidth]{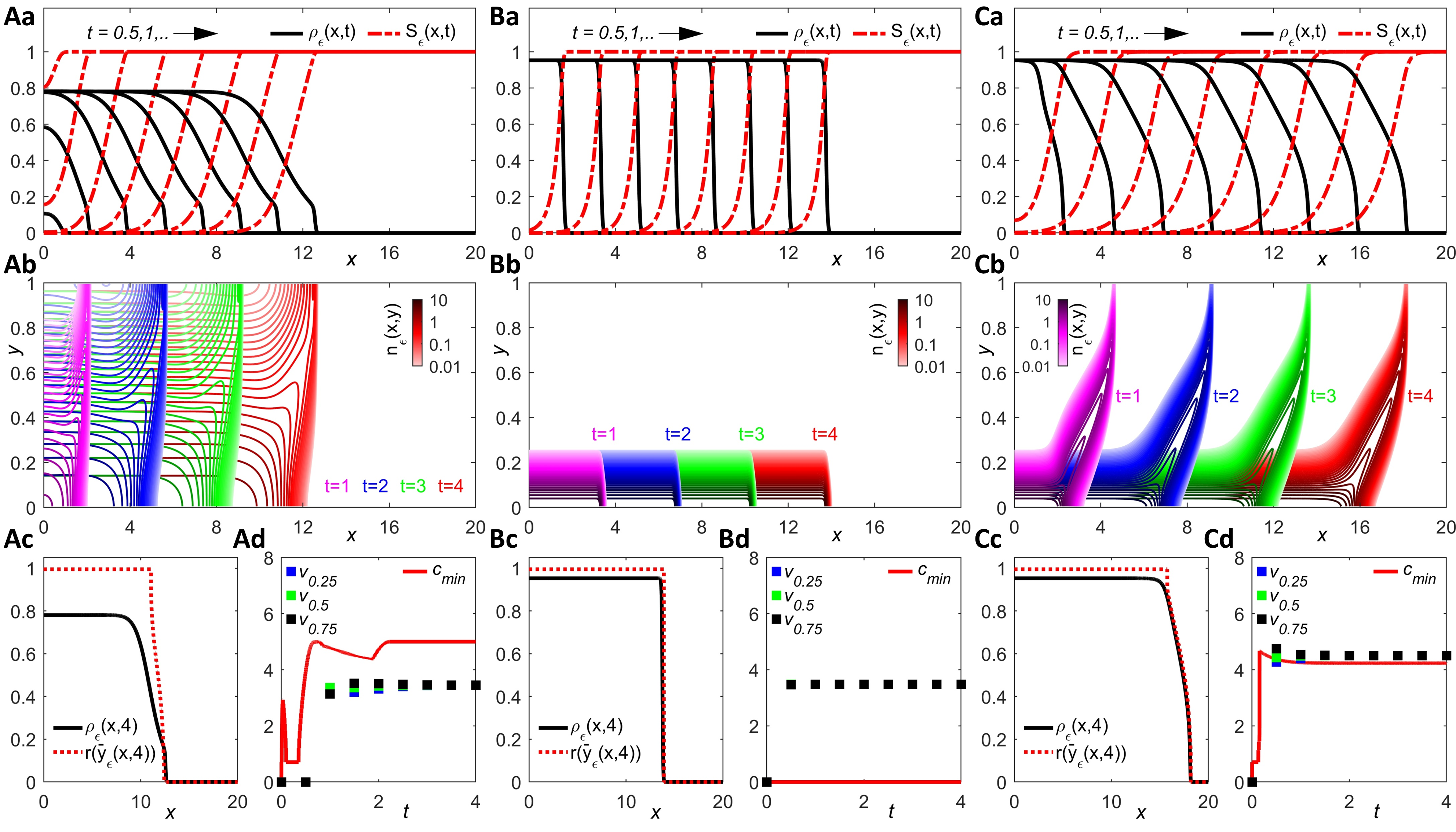}
\end{center}
\caption{{\bf Evolution to travelling-wave solutions under certain other scenarios.} Model formulation and parameter values as in Figure~\ref{fig:figure3}A, except {\bf A} $\e = 0.1$ (rather than $\e = 0.01$), {\bf B} $\alpha = 0$ (rather than $\alpha = 10$), and in {\bf C} an additional attractant diffusion term has been included (i.e. the term $D \partial^2_{xx} S_{\e}$ has been added to the right-hand side of~\eqref{eq:PDESeps}, where we choose $D=1$). {\bf a} Total cell density $\rho_{\e}(x,t)$ (solid black lines) and attractant concentration $S_{\e}(x,t)$ (dashed red lines) plotted at times $t=0,0.5,1,\hdots,4$. {\bf b} Cell population density $n_{\e}(x,y,t)$, where light to dark contour lines indicate increasing density and different tones discriminate the distributions at (magenta) $t=1$, (blue) $t=2$, (green) $t=3$, (red) $t=4$. {\bf c} Comparison between $\rho_{\e}(x,4)$ (solid black) and $r(\bar{y}_{{\e}}(x,4))$ (dotted red). {\bf d} Comparison between front propagation speeds $v_{S_i}$ computed numerically via~\eqref{eq:wfprspS} for three thresholds $S_i$ and the minimal wave speed $c_{{\rm min}}$ computed numerically via the formula~\eqref{eq:minws1}.}\label{fig:figureappendix}
\end{figure}

\end{document}